%% file: main.tex
\documentclass[twocolumn]{aastex701}
\usepackage{amsmath}
\usepackage{graphicx}
\usepackage{subcaption}
\usepackage{placeins}

\DeclareRobustCommand{\okina}{%
  \raisebox{\dimexpr\fontcharht\font`A-\height}{%
    \scalebox{0.8}{`}%
  }%
}

\begin{document}

\title{Early-Excess SNe\,Ia Show Strong \ion{C}{2} Absorption}

\author[0009-0000-4894-7058]{L.~P.~Derose}
\affiliation{Institute for Astronomy, University of Hawai\okina i at Mānoa, 2680 Woodlawn Dr., Honolulu, HI 96822, USA}
\affiliation{Hamilton College, 198 College Hill Rd, Clinton, NY 13323, USA}
\email[show]{lderose@hamilton.edu}

\author[0000-0003-3953-9532]{W.~B.~Hoogendam}
\altaffiliation{NSF Graduate Research Fellow}
\affiliation{Institute for Astronomy, University of Hawai\okina i at Mānoa, 2680 Woodlawn Dr., Honolulu, HI 96822, USA}
\email[show]{willemh@hawaii.edu} 

\author[0000-0003-4631-1149]{B.~J.~Shappee}
\affiliation{Institute for Astronomy, University of Hawai\okina i at Mānoa, 2680 Woodlawn Dr., Honolulu, HI 96822, USA}
\email{shappee@hawaii.edu}

\author[0000-0002-5221-7557]{C.~Ashall}
\affiliation{Institute for Astronomy, University of Hawai\okina i at Mānoa, 2680 Woodlawn Dr., Honolulu, HI 96822, USA}
\email{cashall@hawaii.edu}

\author[0000-0002-6230-0151]{D.~O.~Jones}
\affiliation{Institute for Astronomy, University of Hawai\okina i at Mānoa, 640 N. A\okina ohoku Pl., Hilo, HI 96720, USA}
\email{dojones@hawaii.edu}

\begin{abstract}
Some type Ia supernovae (SNe\,Ia) exhibit an early flux excess in their rising light curves, but the physical origin of this behavior remains uncertain. The processes driving these excesses affect the ejecta at early times and may also imprint on the outermost layers. We investigate whether early excesses are associated with differences in unburned material by measuring pre-maximum carbon absorption. Using multi-component Gaussian fits to the blended \ion{Si}{2} $\lambda6355$ and \ion{C}{2} $\lambda6580$ region, we measure pseudo-equivalent widths and velocities for a sample of 35 nonpeculiar SNe\,Ia with rising light curve categorizations. We find that early-excess objects ($n=13$) tend to exhibit stronger \ion{C}{2} absorption than non-excesses ($n=22$) at phases earlier than approximately $-13$~d, with a two-sample K-S test of the earliest measurements yielding $p=0.025$. In contrast, the HV \ion{Si}{2} components do not clearly separate the two groups. Early-excess SNe\,Ia tend to be more luminous and decline more slowly, but objects with similar maximum-light properties span a range of \ion{C}{2} pseudo-equivalent widths, suggesting that this effect is distinct from the known luminosity-width relation. We show that, at least in some cases, when objects have similar light-curve shape and photospheric temperature, the effect we see is not driven by ionization but by distinct differences in the explosion, such as carbon abundance and density and the efficiency of burning in the WD. Our results provide a new constraint on explosion models for early excesses, as they must produce a tendency for stronger early \ion{C}{2} absorption.

\end{abstract}

\section{Introduction}
Type Ia Supernovae (SNe\,Ia) have long been attributed to thermonuclear explosions of carbon-oxygen white dwarf (CO WD) stars \citep{hoy60, Bloom12}. Not only are they drivers of chemical evolution and the main producers of iron in the universe \citep{Matteucci01}, but their
usefulness as cosmological distance indicators \citep{Phillips93} led to the discovery of the accelerating expansion of the universe \citep{riess98, Perlmutter99}. However, the progenitor scenarios and explosion mechanisms underlying these phenomena remain debated (see, e.g., \citealp{Maoz14, Jha19, Ruiter2025} for reviews).

Proposed SN\,Ia progenitors have traditionally been divided into the single degenerate (SD) and double degenerate (DD) scenarios (see, e.g., \citealp{Maoz14, Ruiter2025} for reviews). In the SD scenario, a WD accretes matter via Roche-Lobe Overflow (RLOF) or stellar wind from a non-degenerate companion star, historically proposed to be a main-sequence, red-giant, or helium star \citep{Whelan73, Nomoto82}. Recent work has increasingly disfavored this system as the dominant progenitor scenario for normal SNe\,Ia, as observational constraints have consistently ruled out red-giant and most main-sequence donors \citep[e.g.,][]{Bianco11, Bloom12, Shappee13a, Tucker20, Do21}. In the DD scenario, angular momentum between at least two CO WDs or a CO WD and a He WD \citep{Perets2019} is gradually lost through common envelope evolution, gravitational wave emission, or both, eventually resulting in an explosion from mass transfer \citep{Piersanti03} or a merger \citep{Webbink84, iben84}. A collision between two WDs has also been proposed to be the origin of some SNe\,Ia \citep{Rosswog09, Raskin09, Thompson11, Katz12, Kushnir13, Shappee13c}. Regardless of the configuration, each channel ultimately leads to a thermonuclear runaway within the primary WD. 

Historically, SNe\,Ia were thought to occur when a WD approached the Chandrasekhar mass ($M_{\text{Ch}} \sim 1.4 \text{M}_{\odot}$), a theoretical limit on the maximum mass that can be supported by electron degeneracy pressure \citep{Chandrasekhar31}.~However, theoretical and observational studies have suggested thermonuclear explosions may also occur in sub-$M_{\text{Ch}}$ and super-$M_{\text{Ch}}$ systems (see, e.g., \citealp{Liu23, Ruiter2025} for reviews), implying that multiple physical pathways to WD disruption may be possible. 

Various explosion mechanisms have been proposed to explain how thermonuclear burning could disrupt a WD. In near-$M_{\text{Ch}}$ explosion models, mass accretion in the SD or DD scenario increases the central density of the WD, eventually leading to carbon ignition. The burning is generally expected to begin as a subsonic deflagration that later transitions into a supersonic detonation (delayed-detonation; \citealp{Khokhlov91, Gamezo04, Blondin13}), although pure deflagrations \citep{Ropke07a} and gravitationally confined detonations \citep{Plewa04} have also been explored. Alternatively, the ignition of a surface He-shell can trigger a secondary detonation in the core of the WD or an edge-lit detonation at sub-$M_{\text{Ch}}$ (double-detonation; \citealp{Nomoto82, Woosley94, Livne95, Fink10, Moll13, Shen14b}). Other DD channels, such as violent mergers and collisions, can produce explosions with masses less than, near, or exceeding $M_{\text{Ch}}$ \citep{iben84, Webbink84, vanKerkwijk10, Scalzo10, Kushnir13}. 

Different explosion pathways can be evaluated by leveraging observations obtained during the earliest phases after the SN. In recent years, untargeted surveys such as the All-Sky Automated Survey for Supernovae (ASAS-SN; \citealp{Shappee14, Kochanek17}), Asteroid Terrestrial-impact Last Alert System (ATLAS; \citealp{Tonry18}), Zwicky Transient Facility (ZTF; \citealp{Bellm19}), and Young Supernova Experiment (YSE; \citealp{Jones21}) have increased the number of SNe\,Ia detected within days after the explosion. While most SNe\,Ia exhibit relative homogeneity near maximum light and at later phases, these early-time observations have revealed substantial photometric and spectroscopic diversity (e.g., \citealp{Liu16, Piro16, Ashall22, Galbany25, Hoogendam25a}). Previous works have used these early signatures to distinguish between competing progenitor scenarios and explosion mechanisms (e.g., \citealp{Nugent11, Cao15, Sai22, Hoogendam24, Ni22, Ni23, Ni25}) and to search for cosmological biases \citep{Ye24}, as differences between model predictions are expected to be particularly pronounced at these epochs. 

One particularly well-studied example of this early-time diversity is the presence of an excess of flux during the rising light curves of some SNe\,Ia (e.g., LSQ12gdj, SNe\,2012cg, 2012fr, 2013dy, 2014J, 2017cbv, ASASSN-18bt, 2021aefx, and 2023bee; \citealp{Zheng13, Scalzo14, Goobar15, Marion16, Hosseinzadeh17, Shappee18, Contreras18, Shappee19, Hosseinzadeh22, Hosseinzadeh23, Wang24}). The light curves of these objects cannot be represented with a single power law, but remain entirely monotonic in their rise (we adopt the \citealp{Hoogendam24} classification scheme where ``bump'' SNe\,Ia must show a non-monotonic light curve). Although their origin remains uncertain, early excesses appear in a non-negligible fraction of SNe\,Ia, and their occurrence rate remains poorly constrained because these short-lived features require sufficiently early and high-cadence observations to detect \citep{Deckers2022, Magee22b}. 

Early-excess SNe\,Ia exhibit varied color evolutions, which may reflect differences in physical origin. They are most commonly observed as UV/blue emission \citep{Stritzinger18, Bulla20}, although the early-time color evolution spans a range of behaviors. A smaller number of events have instead claimed red early-time colors or redward evolution \citep[e.g.,][]{Jiang17, Ni22, Pearson24}. Recently, \citet{Wu25} found that early-excess normal SNe\,Ia commonly exhibit either a small red bump or a blue plateau in their early $B-V$ evolution.

Several models have been proposed to explain early-time flux excesses, including interaction between supernova ejecta and a non-degenerate companion \citep{Kasen10}, shock interaction with circumstellar material (CSM; \citealp{Levanon15, Piro16}), the presence of $^{56}$Ni near the surface \citep{Piro13, Piro16, Noebauer17, Magee20b}, and helium-shell double-detonations \citep{Shen18, Polin19, Magee21}. In addition to generating extra flux, each model should also imprint the outer ejecta, making spectral observations potentially  useful for investigating the origin of early excesses.

Early-time spectra are particularly valuable because observations obtained before maximum light primarily probe the outer layers, where unburned material is most likely to reside \citep{Thomas07, Nugent11, Silverman12C}.~The composition and distribution of this material can constrain SN\,Ia progenitor systems and explosion mechanisms \citep{Parrent2011, Heringer2019}. Both carbon and oxygen trace unburned material, but oxygen is a less reliable diagnostic because it may be produced during carbon burning \citep{Thomas11, Folatelli12, Zhao16}. Carbon therefore provides a clean observational test of explosion models, which predict different distributions of unburned material \citep{Iwamoto99, Parrent2011, Heringer2019}. For example, delayed-detonation models generally undergo more complete burning than pure deflagrations \citep{Gamezo04, Ma13}. Sub-$M_{\text{Ch}}$ double-detonation models may yield substantial intermediate-mass elements while leaving little unburned carbon \citep{Fink10, Polin19}, while a merger of two WDs may produce observable carbon through a C-rich envelope, ejecta mixing, or asymmetries in the explosion \citep[e.g.,][]{Raskin13, Moll14, Raskin14}. 

In the optical regime, \ion{C}{2}~$\lambda6580$ is the most direct observational signature of unburned material. Carbon in the outer ejecta of normal SNe\,Ia is expected to be predominantly singly ionized \citep{Tanaka2008} and \ion{C}{2}~$\lambda6580$ is the strongest optical transition of \ion{C}{2} \citep{Hatano1999}. Its observed strength and detectability depend on the extent of burning, ionization state, and ejecta asymmetry/viewing angle \citep{Thomas07, Tanaka2008, Scalzo10}.~The feature is frequently weak and blended with the neighboring \ion{Si}{2} $\lambda6355$ P-Cygni profile, making secure identifications difficult \citep{Salvo01, Stanishev07, Parrent2011}. Previous studies have detected \ion{C}{2} absorption in approximately 20--40\% of SNe\,Ia observed at least five days before maximum light, although this rate depends on data quality, spectral coverage, methodology, and the availability of sufficiently early observations \citep{Thomas11, Folatelli12, Silverman12C}. 

Previous work linking carbon to photometric properties has focused on light-cruve decline rate, peak luminosity, and color. \citet{Thomas11} and \citet{Maguire2014} found that SNe\,Ia with carbon-positive spectra tend to have faster-declining light curves, although \citet{Silverman12C} found no such relationship. Carbon has been used as a valuable probe of explosion physics, but apart from individual objects such as the underluminous SN\,2022xkq, which showed strong \ion{C}{2} and \ion{C}{1} absorption with a red flux excess \citep{Pearson24}, the connection between carbon and the mechanisms producing early-time flux excesses remains unexplored at the population level.

In this work, we investigate whether SNe\,Ia exhibiting early excesses differ systematically in the detection or strength of their \ion{C}{2}~$\lambda6580$ absorption. \S\ref{sec:sample} describes our sample selection and observational constraints. \S\ref{sec:methods} outlines the methodology used to measure carbon absorption while decomposing the broader silicon profile, including HV \ion{Si}{2}, and estimating uncertainties. The resulting measurements are presented in \S\ref{sec:results}. In \S\ref{sec:discussion}, we discuss the implications of our findings for the origin of flux excesses and the broader SNe\,Ia progenitor systems and explosion mechanisms. Finally, our conclusions are summarized in \S\ref{sec:conclusion}. Throughout this manuscript we assume a flat $\Lambda$CDM cosmology with $H_0 = 70~\mathrm{km~s^{-1}~Mpc^{-1}}$ and $\Omega_M = 0.3$. 

\section{Sample Selection} \label{sec:sample}
We compile a sample of SNe\,Ia with available early-time optical spectroscopy and published flux excess classifications. We use the list of SNe\,Ia with categorized rises from \citet{Hoogendam24}, supplemented by SN\,2023wrk \citep{Liu2024} and SN\,2024epr \citep{Hoogendam25a}, which were not available when \citet{Hoogendam24} was published.~We omit SNe\,Iax because they represent a distinct physical class from normal SNe\,Ia \citep{li03, Phillips07, Foley13, Stritzinger14, Stritzinger15, Magee16, Tomasella16, Jha17, Maguire23, Hoogendam25_24pxl, Magee25}. We also exclude 2002es-like and 2003fg-like SNe\,Ia, which are the only subtypes known to exhibit early-time non-monotonic bumps \citep{Hoogendam24}, likely distinct from the early excesses investigated here. As with SNe\,Iax, there is strong evidence that these subtypes do not arise from the same physical origins as normal SNe\,Ia \citep{Chen19, Hsiao20, Ashall21, Lu21, Siebert24, Kwok24, Bose25, Paniagua26}. We remove SN\,2014J because the high host-galaxy extinction makes continuum placement and carbon absorption measurements unreliable \citep{Ashall14}, and SN\,2020tld because it lacks sufficient photometric coverage in the $B$-band near maximum light, preventing a reliable phase estimate. From the initial set of 44 candidates, this yields a final sample of 35 SNe\,Ia. Table~\ref{tab:sample} summarizes the object properties compiled from the literature, including rising light curve categorization, redshift, MJD date of $B$-band maximum, $\Delta m_{15}(B)$, $M_B$, and subtype. For some sources, the original works do not provide uncertainties on their measurements of $\Delta m_{15}(B)$ and $M_B$. In these cases, we estimate them from published photometric, extinction, and distance uncertainties. 

\begin{figure}[htbp]
\centering
\includegraphics[width=\linewidth]{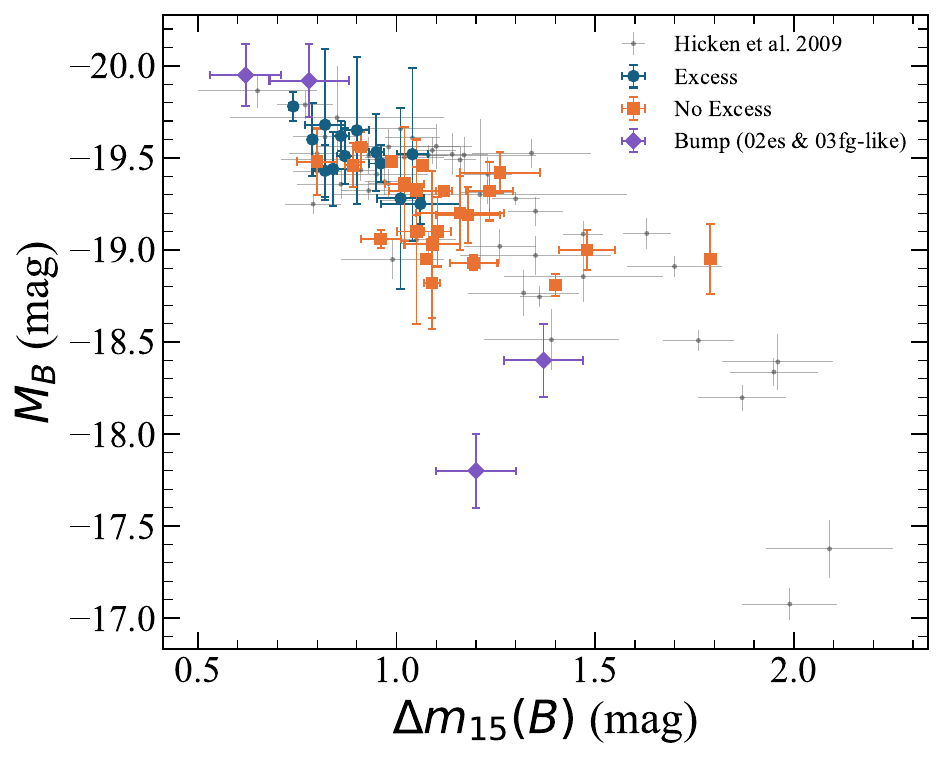}
\caption{Luminosity-width relation for our sample and reference SNe\,Ia from \citet{Hicken09} shown in gray. Blue, orange, and purple points indicate objects with early-time flux excesses, no detected excesses, and non-monotonic early bumps, respectively. Classifications follow \citet{Hoogendam24}.}
\label{fig:lwr}
\end{figure}
Figure~\ref{fig:lwr} shows our sample on the luminosity-width relationship (LWR; \citealp{Phillips93}) using objects from \citet{Hicken09} for comparison. We restrict the comparison sample to objects with $z_{\rm CMB} > 0.015$ to reduce the impact of peculiar velocities on redshift-based distance estimates. For context, we also include four 2002es-like and 2003fg-like SNe\,Ia. Our selected SNe\,Ia span the range of normal $\Delta m_{15}(B)$ and $M_B$ values, although those with flux excesses tend to have broader light curves and brighter peak magnitudes, consistent with the findings of \citet{Jiang18}.

\begin{table*}[ht]
\centering
\caption{Object information for the SNe\,Ia sample. Unless otherwise noted,
rising light-curve classifications are adopted from \citet{Hoogendam24}.
Heliocentric host-galaxy redshifts are adopted from the references indicated
by superscript letters. The date of $B$-band maximum, $\Delta m_{15}(B)$,
$M_B$, and subtype are listed for each object. Uncertainties marked with $^{*}$ were not explicitly reported; we estimate them from the published photometric, extinction, and distance uncertainties.}
\label{tab:sample}

\footnotesize
\setlength{\tabcolsep}{5pt}
\renewcommand{\arraystretch}{1.05}

\begin{tabular}{lcccccc}
\hline
Object &
Flux Excess &
Redshift ($z$) &
$MJD_{B,\mathrm{max}}$ &
$\Delta m_{15}(B)$ (mag) &
$M_B$ (mag) &
Subtype \\
\hline
SN\,2009ig & No & $0.0088^{a}$ & 55080.0 (1) & $0.89 \pm 0.02$ (1) & $-19.46 \pm 0.12$ (32) & Norm \\
SN\,2011fe & No & $0.0008^{b}$ & 55814.5 (2) & $1.10 \pm 0.04$ (2) & $-19.10 \pm 0.19$ (2) & Norm \\
SN\,2012cg & Yes & $0.001458^{c}$ & 56081.3 (4) & $0.86 \pm 0.02$ (4) & $-19.62 \pm 0.08$ (4) & Norm \\
SN\,2012fr & Yes & $0.005457^{d}$ & 56242.6 (5) & $0.82 \pm 0.03$ (5) & $-19.43 \pm 0.14$ (5) & Norm \\
LSQ12gdj & Yes & $0.030324^{g}$ & 56252.5 (3) & 
$0.74 \pm 0.01$ (3) & $-19.78 \pm 0.08^{*}$ (3) & 91T \\
SN\,2012ht & No & $0.00356^{e}$ & 56295.6 (6) & $1.19 \pm 0.06$ (27) & $-18.93 \pm 0.04^*$ (35) & Fast \\
iPTF13dge & No & $0.01586^{i}$ & 56558.0 (7) & $1.18 \pm 0.08^{*}$ (34) & $-19.19 \pm 0.15^{*}$ (34) & Norm \\
SN\,2013dy & Yes & $0.003889^{f}$ & 56500.7 (9) & $0.90 \pm 0.03$ (45) & $-19.65 \pm 0.40$ (45) & Norm \\
iPTF13ebh & No & $0.01316^{j}$ & 56622.9 (8) & $1.79 \pm 0.01$ (8) & $-18.95 \pm 0.19$ (8) & Fast \\
SN\,2013gy & No & $0.014023^{h}$ & 56648.5 (10) & $1.23 \pm 0.06$ (10) & $-19.32 \pm 0.16$ (10) & Norm \\
iPTF14bdn & Yes & $0.01558^{k}$ & 56822.5 (12) & $0.84 \pm 0.05$ (12) & $-19.44 \pm 0.15$ (12) & 99aa \\
ASASSN-14lp & No & $0.0051^{a}$ & 57015.3 (11) & $0.80 \pm 0.05$ (11) & $-19.48 \pm 0.18^{*}$ (34) & Norm \\
SN\,2015bq & Yes & $0.02804^{k}$ & 57084.1 (13) & $0.82 \pm 0.05$ (13) & $-19.68 \pm 0.41$ (13) & 99aa \\
SN\,2015F & No & $0.00489^{a}$ & 57106.3 (51) & $1.26 \pm 0.10$ (51) & $-19.42 \pm 0.11$ (51) & Fast \\
iPTF16abc & No & $0.02328^{l}$ & 57498.8 (14) & $0.91 \pm 0.02$ (36) & $-19.56 \pm 0.08$ (34) & 99aa \\
SN\,2017cbv & Yes & $0.004^{m}$ & 57840.4 (15) & $1.06 \pm 0.10^{*}$ (48)$^{\dagger}$ & $-19.25 \pm 0.11^{*}$ (48, 46)$^{\dagger}$ & Norm \\
SN\,2017cfd & No & $0.01209^{i}$ & 57843.4 (16) & $1.16 \pm 0.11$ (16) & $-19.20 \pm 0.20$ (16) & Norm \\
SN\,2017cyy & No & $0.00978^{a}$ & 57870.1 (17) & $1.07 \pm 0.01$ (17) & $-19.46 \pm 0.03$ (17) & Norm \\
SN\,2017erp & No & $0.006174^{n}$ & 57934.4 (17) & $0.99 \pm 0.01$ (17) & $-19.48 \pm 0.03$ (17) & Norm \\
SN\,2017fgc & No & $0.007722^{o}$ & 57959.5 (18) & $1.05 \pm 0.07$ (39) & $-19.32 \pm 0.13$ (39) & Norm \\
SN\,2018gv & No & $0.00527^{g}$ & 58149.7 (22) & $0.96 \pm 0.05$ (22) & $-19.06 \pm 0.05$ (22) & Norm \\
ASASSN-18bt & Yes & $0.01098^{p}$ & 58162.7 (19) & $0.96 \pm 0.03$ (19) & $-19.47 \pm 0.10$ (19) & Norm \\
SN\,2018xx & No & $0.00999^{q}$ & 58183.9 (17) & $1.48 \pm 0.07$ (44) & $-19.00 \pm 0.11$ (44) & Norm \\
SN\,2018yu & No & $0.008112^{b}$ & 58194.3 (17) & $1.05 \pm 0.05$ (40) & $-19.10 \pm 0.50$ (40) & Norm \\
SN\,2018agk & No & $0.02613^{r}$ & 58204.2 (20) & $1.07 \pm 0.01$ (20) & $-18.95 \pm 0.01$ (20) & Norm \\
SN\,2018aoz & No & $0.0058^{s}$ & 58222.2 (21) & $1.12 \pm 0.02$ (21) & $-19.32 \pm 0.01$ (21) & Norm \\
SN\,2019np & Yes & $0.00452^{u}$ & 58510.2 (24) & $1.04 \pm 0.04$ (24) & $-19.52 \pm 0.47$ (24) & Norm \\
SN\,2019ein & No & $0.00775^{v}$ & 58619.5 (23) & $1.40 \pm 0.01$ (23) & $-18.810 \pm 0.059$ (23) & Norm \\
SN\,2020nlb & No & $0.00243^{q}$ & 59041.8 (25) & $1.09 \pm 0.02$ (50)$^{\ddagger}$ & $-18.82 \pm 0.25$ (50)$^{\ddagger}$ & Norm \\
SN\,2021fxy & No & $0.0094^{w}$ & 59305.2 (49) & $1.02 \pm 0.05$ (49) & $-19.36 \pm 0.31$ (49) & Norm \\
SN\,2021hpr & Yes & $0.009346^{x}$ & 59321.4 (28) & $0.95 \pm 0.02$ (42) & $-19.53 \pm 0.21$ (42) & Norm \\
SN\,2021aefx & Yes & $0.005017^{t}$ & 59546.5 (26) & $1.01 \pm 0.06$ (41) & $-19.28 \pm 0.49$ (41) & Norm \\
SN\,2023bee & Yes & $0.0067^{i}$ & 59992.3 (29) & $0.79 \pm 0.01$ (43) & $-19.60 \pm 0.20$ (43) & Norm \\
SN\,2023wrk & Yes (30) & $0.0230^{y}$ & 60269.5 (30) & $0.87 \pm 0.02$ (30) & $-19.51 \pm 0.15$ (30) & 91T/99aa \\
SN\,2024epr & No (31) & $0.0053^{z}$ & 60406.6 (31) & $1.09 \pm 0.07$ (31) & $-19.03 \pm 0.40$ (31) & Norm \\
\hline
\end{tabular}

\vspace{0.4em}
\begin{minipage}{0.98\textwidth}
\scriptsize
(1) \citet{Foley12}; (2) \citet{Pereira13}; (3) \citet{Scalzo14};
(4) \citet{Marion16}; (5) \citet{Contreras18}; (6) \citet{Yamanaka14};
(7) \citet{Ferretti16}; (8) \citet{Hsiao15}; (9) \citet{Zheng13};
(10) \citet{Holmbo19}; (11) \citet{Shappee16}; (12) \citet{Smitka15};
(13) \citet{Li22}; (14) \citet{Ferretti17}; (15) \citet{Wee18};
(16) \citet{Han20}; (17) \citet{Burke22}; (18) \citet{Burke21};
(19) \citet{Li19}; (20) \citet{Wang21}; (21) \citet{Ni22};
(22) \citet{Yang20}; (23) \citet{Pellegrino20}; (24) \citet{Sai22};
(25) \citet{Sand21}; (26) \citet{Hosseinzadeh22}; (27) \citet{Morrell24}; (28) \citet{Ward23};
(29) \citet{Hosseinzadeh23}; (30) \citet{Liu2024};
(31) \citet{Hoogendam25a}; (32) \citet{Marion13};
(33) \citet{Childress13}; (34) \citet{Jiang18}; (35) \citet{uddin24};
(36) \citet{Dhawan18}; (37) \citet{Cartier17};
(39) \citet{Zeng21}; (40) \citet{Graham22}; (41) \citet{Ashall22};
(42) \citet{Zhang22}; (43) \citet{Wang24}; (44) \citet{Li23}; (45) \citet{Zhai16}; (46) \citet{Sand18};
(47) \citet{Burns14}; (48) \citet{Hosseinzadeh17}; (49) \citet{Li26}; (50) \citet{Izzo26}; (51) \citet{Im15}.

\medskip
$^{a}$ \citet{Meyer04}; $^{b}$ \citet{deVaucouleurs91};
$^{c}$ \citet{Kent08}; $^{d}$ \citet{Bureau96};
$^{e}$ \citet{Kerr86}; $^{f}$ \citet{Schneider92};
$^{g}$ \citet{Springob05}; $^{h}$ \citet{Catinella05};
$^{i}$ \citet{Falco99}; $^{j}$ \citet{vdBosch15};
$^{k}$ \citet{Albareti17}; $^{l}$ \citet{Beers95};
$^{m}$ \citet{Koribalski04}; $^{n}$ \citet{Theureau05};
$^{o}$ \citet{Cappellari11}; $^{p}$ \citet{Schneider90};
$^{q}$ \citet{Smith2000}; $^{r}$ \citet{Jones09};
$^{s}$ \citet{Norris11}; $^{t}$ \citet{Bilicki14};
$^{u}$ \citet{Rhee96}; $^{v}$ \citet{vanDriel01};
$^{w}$ \citet{Rothberg06}; $^{x}$ \citet{Epinat08};
$^{y}$ \citet{Liu2024}; $^{z}$ \citet{Huchra99}.

\medskip
$^{\dagger}$ Other published estimates of $\Delta m_{15}(B)$ for SN\,2017cbv include $0.88\pm0.07$ mag from \citet{Wee18} and $0.99\pm0.01$ mag from \citet{Wang20}, although we adopt $1.06\pm0.10^*$ mag from \citet{Hosseinzadeh17}. We use the improved distance estimate from \citet{Sand18} to calculate the absolute magnitude.

$^{\ddagger}$ Other published estimates of $\Delta m_{15}(B)$ for SN\,2020nlb include $1.29\pm0.06$ mag from \citet{Sand21} and $1.28\pm0.02$ mag from \citet{Williams24}, although we adopt the SNooPy value of $1.089\pm0.018$ mag from \citet{Izzo26}. For the absolute magnitude, we use the $B$-band peak magnitude reported by \citet{Izzo26} together with their adopted independent host-galaxy distance. Previous estimates include $M_B=-19.25$ mag from \citet{Sand21} and $M_B=-19.1\pm0.2$ mag from \citet{Williams24}.
\end{minipage}
\end{table*}

For the 35 SNe\,Ia in the final sample, we obtain spectra from WISeREP\footnote{\url{https://www.wiserep.org/}} \citep{Yaron2012} or directly from the corresponding authors when not publicly available.~Our spectral dataset strictly includes optical spectra taken before $B$-band maximum. We exclude individual spectra whose wavelength coverage does not encompass the \ion{C}{2} $\lambda6580$ fitting region and both adjacent pseudo-continuum windows (see \S\ref{sec:methods}) and those with a low signal-to-noise ratio (SNR). These exclusions remove individual spectra but do not change the final object sample.

\section{Methods} \label{sec:methods}
\ion{C}{2} $\lambda6580$ lies in a crowded spectral region, so identifying the presence of carbon is largely dependent on distinguishing signal from noise and the effects of line blending with neighboring features \citep{Parrent2011}. The dominant source of blending is the prominent \ion{Si}{2} $\lambda6355$, which may include both a high-velocity (HV) and a photospheric component. HV \ion{Si}{2} is common in early SN\,Ia spectra and is estimated to be present in roughly three-quarters of spectra before $-11$~d \citep{Harvey25}. Since \ion{C}{2} $\lambda6580$ can overlap with the photospheric \ion{Si}{2} component, we model this region using a composite Gaussian profile following previous works \citep{Parrent2011, Zhao2021}. This approach lets us isolate absorption from carbon while also measuring the \ion{Si}{2} components.

Our analysis is based on the Measure Intricate Spectral Features In Transient Spectra (\texttt{misfits}; \citealp{Holmbo23})\footnote{\url{http://github.com/sholmbo/misfits}} and Spextractor \citep{Burrow20}\footnote{\url{https://anthonyburrow.github.io/projects/Spextractor/}} packages. For each spectrum, we define a pseudo-continuum between two local maxima in the rest frame. We manually select two wavelength windows of approximately $25$ \AA\ and fit a straight line through the mean flux in each window. The resulting continuum normalizes the spectrum between the two continuum endpoints. We then use SciPy's \texttt{curve\_fit} with weighted least squares to model the normalized absorption profile as the sum of Gaussian components.

We select initial parameter estimates interactively using the observed profile and the expected locations of the relevant features. We typically initialize the widths of the photospheric \ion{Si}{2}, HV \ion{Si}{2}, and \ion{C}{2} components to 150, 80, and 50 \AA\, respectively, and adjust these values to match the shape of the absorption profile. These initial values guide the optimizer; the fit determines the final parameters. This process does not eliminate degeneracies in strongly blended features, although we visually inspect each fit and reject solutions with implausible widths or component centers inconsistent with the expected velocities. We convert the central wavelength of each fitted component to a velocity using its corresponding rest wavelength and calculate the pseudo-equivalent width (pEW; \citealp{Garavini07}) from the Gaussian area:
\begin{equation}
pEW_i = d_i \sigma_i\sqrt{2\pi}.
\end{equation}

To determine the preferred Gaussian components for each spectrum, we compare models with different combinations using the Bayesian Information Criterion (BIC). The BIC penalizes models that improve the fit by adding unnecessary parameters. Thus, we adopt the lowest-BIC model when it is also consistent with the expected velocity evolution of the features and with the broader spectral time series. This prevents isolated noisy features or unrelated line blending from being interpreted as \ion{C}{2} solely because they improve the fit to a single spectrum.

\begin{figure*}[htbp]
    \centering
    \includegraphics[width=\textwidth]{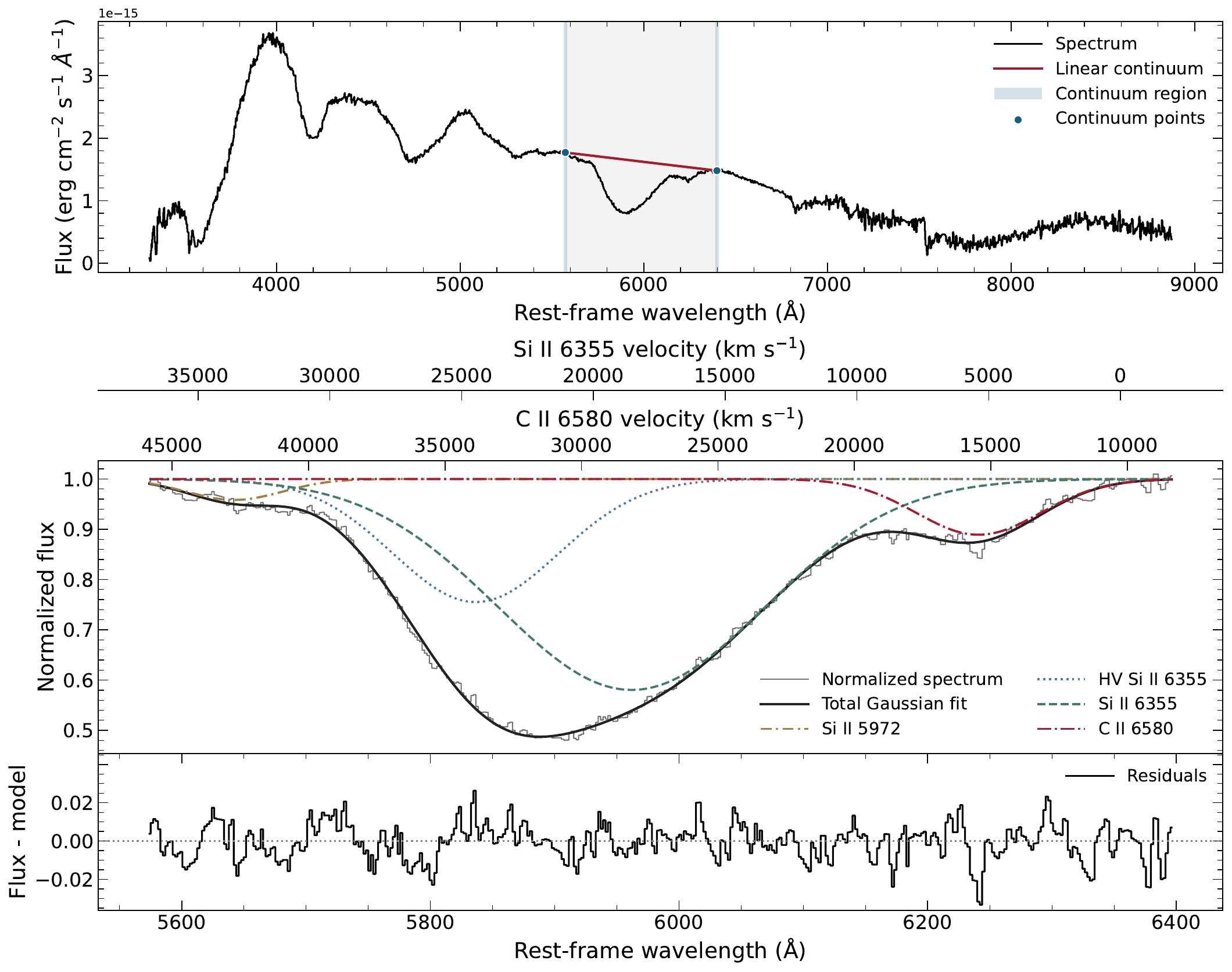}
    \caption{Example of a typical spectral decomposition; SN\,2023bee is shown at $-14.1$~d. Continuum regions with mean endpoints and fitted line (top). Normalized spectrum with total fit and each Gaussian component representing \ion{Si}{2} $\lambda5972$, HV and photospheric \ion{Si}{2} $\lambda6355$, and \ion{C}{2} $\lambda6580$ (middle). Model residuals from the normalized spectrum (bottom).}
    \label{fig:fit2023bee}
\end{figure*}

Broad absorption from \ion{Si}{2} $\lambda5972$ an also affect the \ion{Si}{2} $\lambda6355$ profile (ASASSN-14lp, SNe\,2009ig, 2012cg, 2015F, 2017fgc, 2018xx, 2019ein, 2020nlb, 2021aefx, 2021fxy, 2021hpr, 2023bee, and 2024epr). In these cases, we expand the model to include an additional Gaussian component when it improves the fit according to the model-selection criteria and better represents the observed blending between the \ion{Si}{2} and \ion{C}{2} features. Figure~\ref {fig:fit2023bee} shows a representative four-Gaussian fit of SN\,2023bee.

In addition to line blending, some spectra are contaminated by interstellar \ion{Na}{1}~D $\lambda\lambda5890,5896$ absorption from the Milky Way and/or host galaxy. If neglected, these narrow features can bias the profile, so we generally mask them prior to fitting. In a few lower-resolution spectra, the \ion{Na}{1}~D doublet is only marginally resolved and appears as a single narrow absorption. Thus, if masking removes too much of the absorption profile or prevents a stable fit, we instead add a narrow Gaussian component to account for the contamination.

To ensure our \ion{C}{2} identifications and measurements are robust, we leverage spectral time series data where available to ensure sensible velocity evolution as a function of time (i.e., as the supernova approaches maximum light, the absorption velocities decrease as the photosphere recedes). We also require the inferred \ion{C}{2} velocity to be broadly consistent with the neighboring \ion{Si}{2} $\lambda6355$ velocity, since previous studies find that they are comparable \citep{Parrent2011, Folatelli12, Silverman12}. This check is not applied as a strict cut, but is used to reject fits whose inferred velocities would be inconsistent with the expected line-forming region. Because \ion{C}{2} $\lambda6580$ is weaker than the neighboring \ion{Si}{2} $\lambda6355$ feature, apparent absorption on the red side of \ion{Si}{2} must be interpreted with caution. For example, in SN\,2021fxy, a redward absorption feature appears near the expected location of \ion{C}{2} in some spectra; however, its evolution is not consistent with the expected behavior of \ion{C}{2} and is instead likely associated with emerging \ion{Fe}{2} $\lambda6456$ closer to maximum light \citep{DerKacy23b}. We attribute absorption redward of \ion{Si}{2} to \ion{C}{2} only when it is consistent with the expected velocity evolution and the broader spectral sequence.

For spectra lacking detectable carbon features, we estimate a $3\sigma$ upper bound from the local root-mean-squared fluctuations around the normalized continuum, the spectral resolution, and the expected width of the \ion{C}{2} feature \citep{Leonard2001, Leonard2007}: 
\begin{equation}
\begin{aligned}
pEW(3\sigma) &= 3\Delta\lambda\Delta I \sqrt{W/\Delta\lambda} \sqrt{1/B} \\
             &= 3\Delta I \sqrt{W\Delta X}
\end{aligned}
 \label{eq:upperlimit}
\end{equation}
where $\Delta\lambda$ is the width of a resolution element, $\Delta I$ is the $1\sigma$ root-mean-square fluctuation of the flux
around a normalized continuum level, $W$ is the FWHM of the expected spectral feature, $B$ is the number of bins per resolution element in the spectrum, and $\Delta X$ is the bin
size of the spectrum.

We estimate the pEW and velocity uncertainties with a Monte Carlo (MC) procedure using $10,000$ realizations. Each trial perturbs the flux at every wavelength according to its uncertainty. To account for continuum-placement uncertainty, each trial independently shifts the manually selected continuum windows by a random value drawn from a uniform distribution between $-50$ and $+50$ \AA\ and recalculates the continuum endpoints from the mean flux within the shifted windows. Then, the MC routine renormalizes and refits the spectrum to obtain new pEWs and velocities. We adopt the 16th, 50th, and 84th percentiles of the resulting distributions as the lower uncertainty, median, and upper uncertainty, respectively.

\begin{figure*}[htbp]
    \centering
    \includegraphics[width=\textwidth]{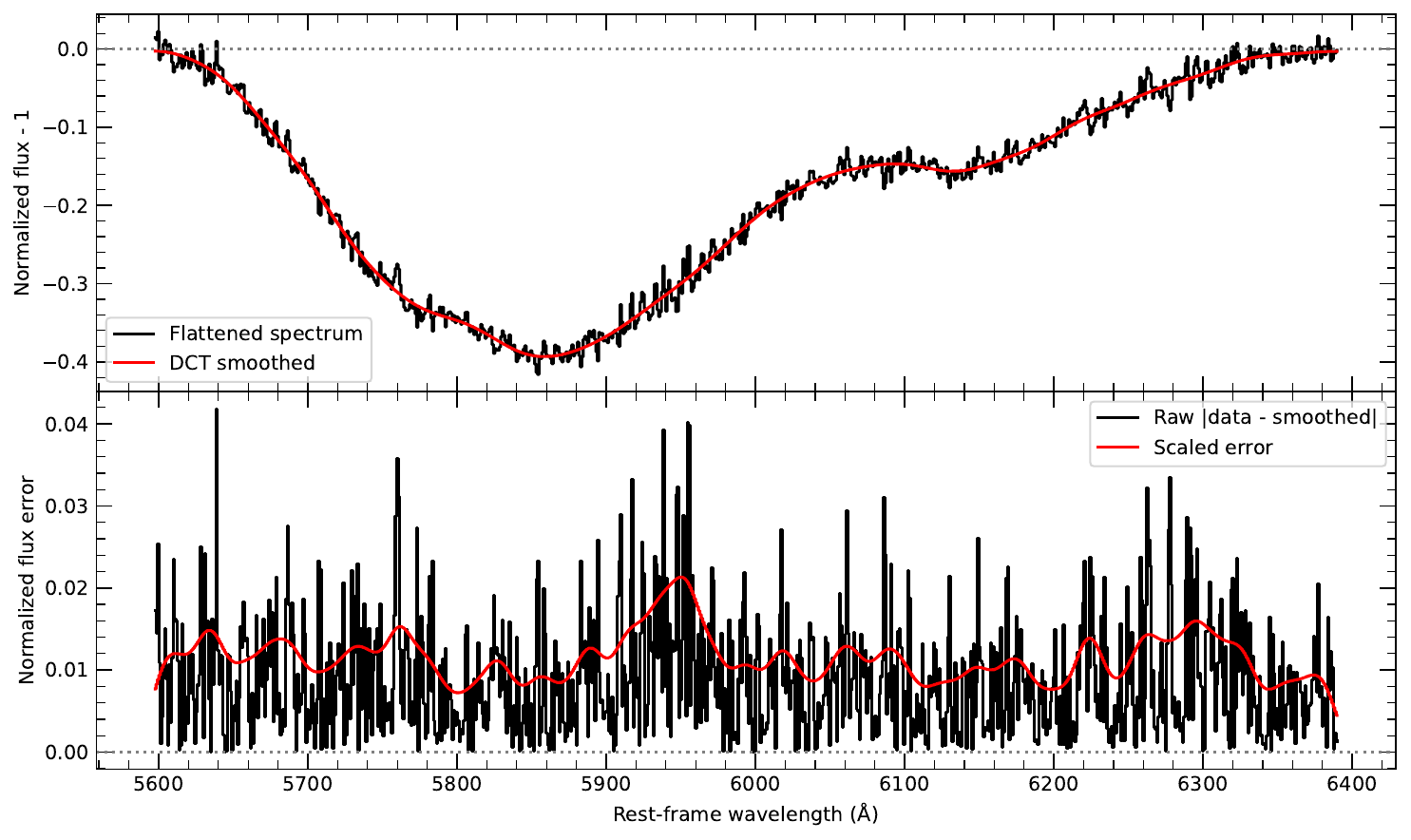}
    \caption{Example of typical error estimation; SN\,2023bee is shown at $-15.3$~d. Flattened spectrum and DCT-smoothed line (top). Raw absolute residuals and the smoothed, rescaled $1\sigma$ uncertainty estimate (bottom).}
    \label{fig:error2023bee}
\end{figure*}

If an uncertainty spectrum is unavailable, we estimate the flux uncertainties using the lowpass rawsmooth technique \citep{Marion09}. We smooth the continuum-normalized spectrum using a Discrete Cosine Transform (DCT),  then smooth the absolute residuals with a Gaussian window and rescale them so that 68\% fall below the error curve. This approximates a local 1$\sigma$ uncertainty; a visualization of error estimation for SN\,2023bee is shown in Figure~\ref{fig:error2023bee}.

\section{Results} \label{sec:results}
\input{tables/measurements_table_preview}
Table~\ref{tab:measurements} lists the measured \ion{C}{2}, \ion{Si}{2}, and HV \ion{Si}{2} pEWs and velocities with their associated uncertainties. In \S\ref{sec:carbon}, we examine whether early-excess SNe\,Ia exhibit differences in the strength or velocity of early-time \ion{C}{2} $\lambda6580$ absorption. In \S\ref{sec:hvsi}, the fitted HV \ion{Si}{2} components are compared as a secondary test of whether any trends are part of broader differences in the structure of the \ion{Si}{2} profile. Finally, in \S\ref{sec:properties}, we compare the $\Delta m_{15}(B)$ and $M_B$ distributions to determine whether the spectroscopic trends could instead reflect differences in light-curve width, peak luminosity, or subtype.

\subsection{Carbon}
\label{sec:carbon}
\begin{figure*}[htbp]
\includegraphics[width=\textwidth]{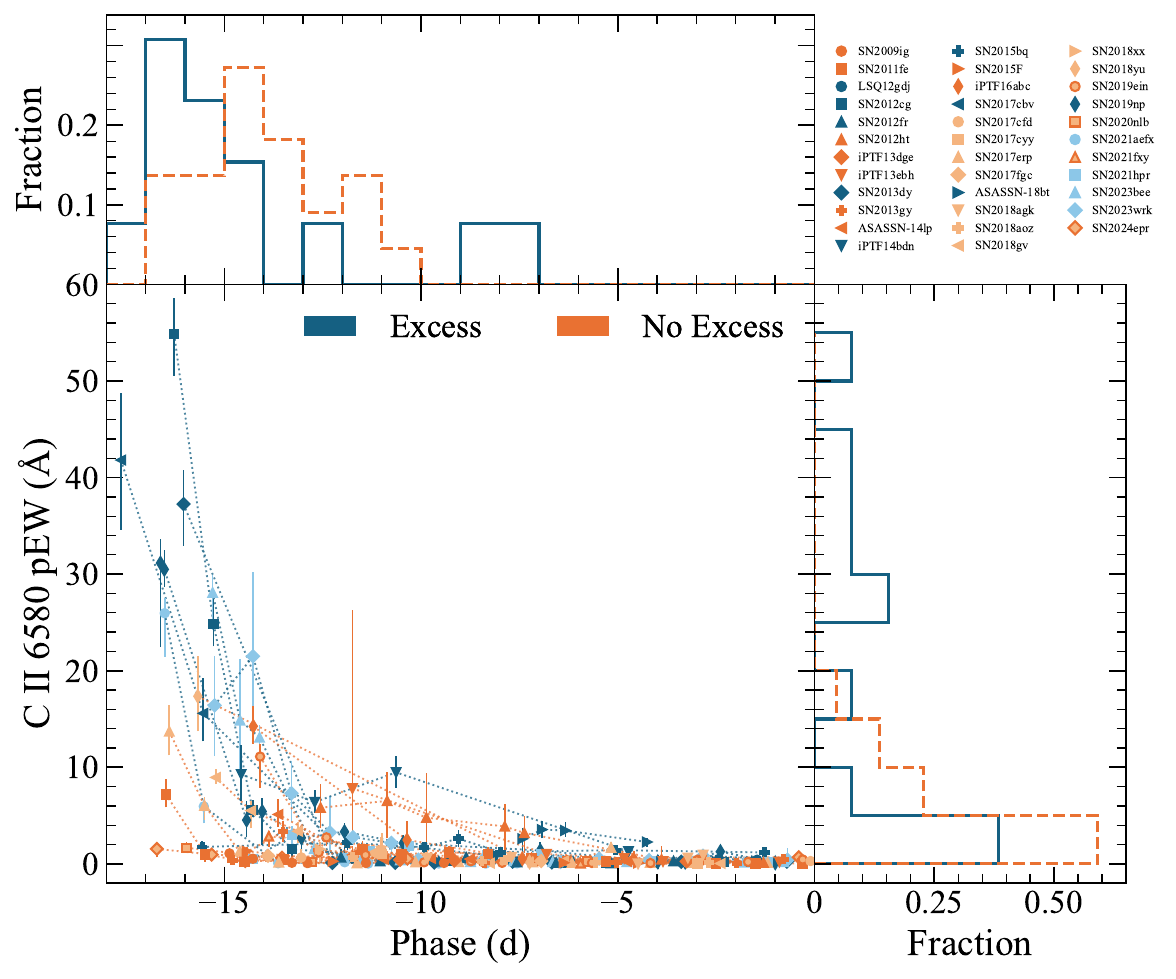}
\caption{\ion{C}{2} $\lambda6580$ pEW measurements of the sample from Table~\ref{tab:measurements}. SNe\,Ia with flux excesses are marked with blue and light blue, and those without are marked with orange and light orange. Different marker shapes and fills identify individual objects, as indicated by the legend. Dotted lines indicate the time evolution of an object. The top histogram shows the phase distribution of the earliest measurement for each object, using 1-day bins. The right histogram shows the distribution of the earliest C pEW measurement for each object, using 5~\AA\ bins.}
\label{fig:cpEW}
\end{figure*}

Figure~\ref{fig:cpEW} shows the evolution of \ion{C}{2} $\lambda6580$ pEW at phases before \(B\)-band maximum. We group the objects according to their rising light-curve categorization, and the histograms show the distributions of the earliest measurements. \ion{C}{2} absorption is detected in at least one spectrum of 28 objects, comprising 10 flux-excess and 18 non-excess SNe\,Ia.

While many objects show some evidence of carbon absorption prior to $-10$~d, the early-excess SNe\,Ia in our sample tend to have larger pEW values than the non-excess objects. In particular, SNe 2012cg, 2013dy, 2017cbv, 2019np, 2021aefx, and 2023bee all have pEWs $>20$ \AA, whereas the largest value among the 22 non-excess objects is $18\pm4$~\AA\ (SN\,2018yu). 

Although the flux-excess objects have slightly earlier phase coverage, the difference in carbon strength remains apparent when comparing overlapping phase ranges. When restricting the phases to $\lesssim-15.5$~d, the four non-excess SNe\,Ia with measured \ion{C}{2} absorption span pEWs from $2\pm1$ to $18\pm4$~\AA,
while SN\,2020nlb has a $3\sigma$ upper limit of
$<2$~\AA. In contrast, five of the six flux-excess objects
span $20\pm1$ to
$55\pm4$~\AA, while SN\,2015bq has a
$3\sigma$ upper limit of $<2$~\AA\ at $-15.6$~d. In addition, from $-15$ to $-14$~d, the six flux-excess objects have a median \ion{C}{2} pEW of 9~\AA, compared with 1~\AA\ for the eight non-excess objects. A weaker separation remains from $-14$ to $-13$~d, after which the distinction becomes less apparent as the carbon feature fades and many measurements become upper limits.

\begin{figure}[htbp]
\centering
\includegraphics[width=\columnwidth]{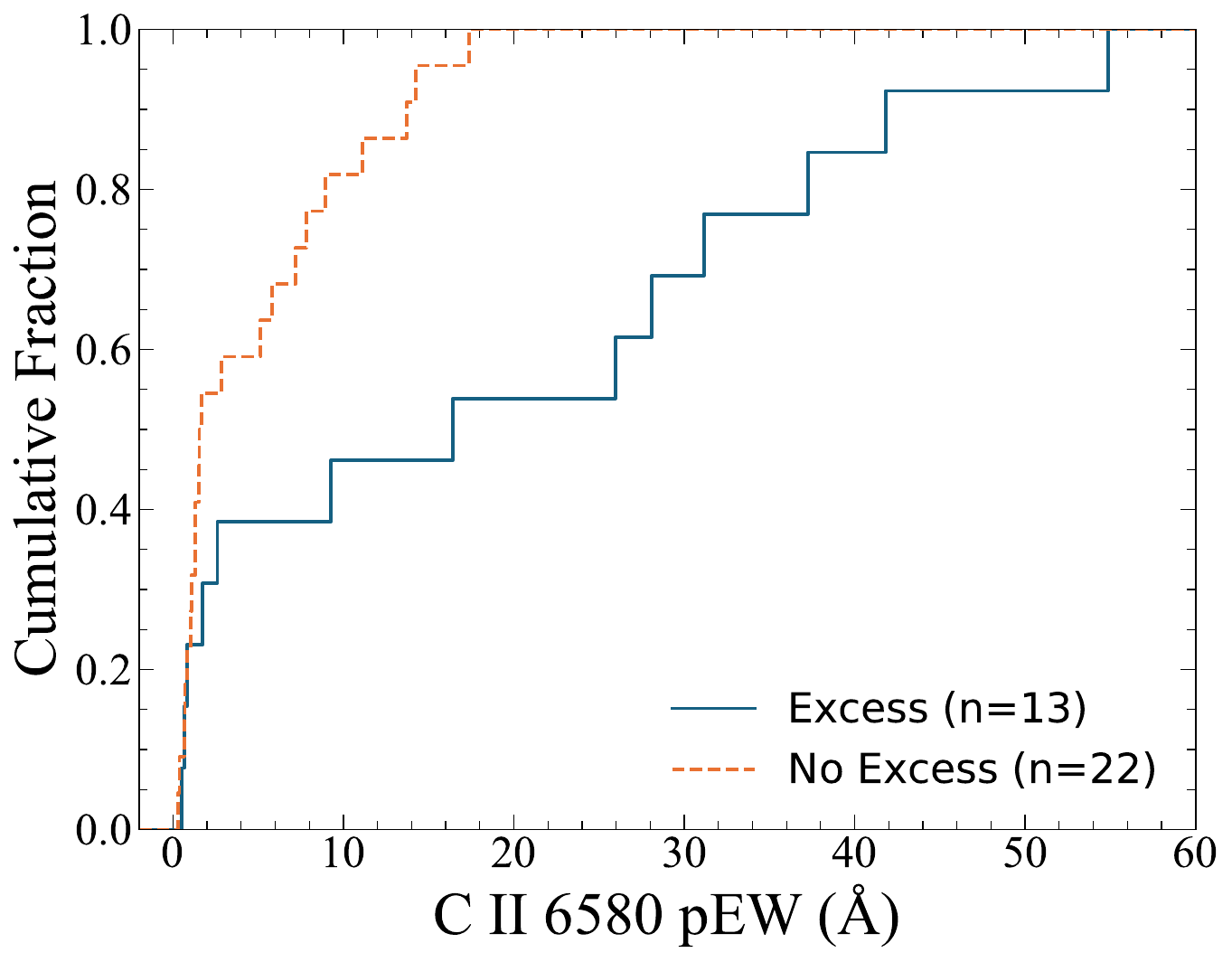}
\caption{Cumulative distributions of the earliest \ion{C}{2} $\lambda6580$ pEW measurements for the flux-excess and non-excess samples. A K--S test yields $D=0.4930$ and $p=0.025$, below the adopted significance level of $\alpha = 0.05$.}
\label{fig:cumulativedist}
\end{figure}

To quantify the difference between the two groups, we compare the cumulative distributions of the earliest \ion{C}{2} $\lambda6580$ pEW measurements in Figure~\ref{fig:cumulativedist}.
Under this convention, a two-sample K--S test yields $D=0.4930$ and $p=0.025$, exceeding $D_{\rm crit}$ at the $\alpha=0.05$ significance level. Thus, the K--S test provides evidence that the two samples differ statistically; early-excess SNe\,Ia tend to show stronger early \ion{C}{2} absorption than non-excess objects. We obtain the same result when setting upper-limit measurements of \ion{C}{2} to zero.

Three of the four non-excess objects with early \ion{C}{2} pEWs above 10~\AA\ (iPTF16abc, SN\,2017erp, and SN\,2018yu) have previously been discussed as possible early-excess events \citep{Jiang18, Burke22}. We retain their non-excess classifications from \citet{Hoogendam24}, since two-component light-curve fits provide only marginal improvements over single-component models. If carbon strength scales with the amplitude or morphology of the early excess rather than with a binary excess/non-excess classification, the non-excess objects with strong carbon may represent intermediate cases.

\begin{figure*}[ht!]
\includegraphics[width=\textwidth]{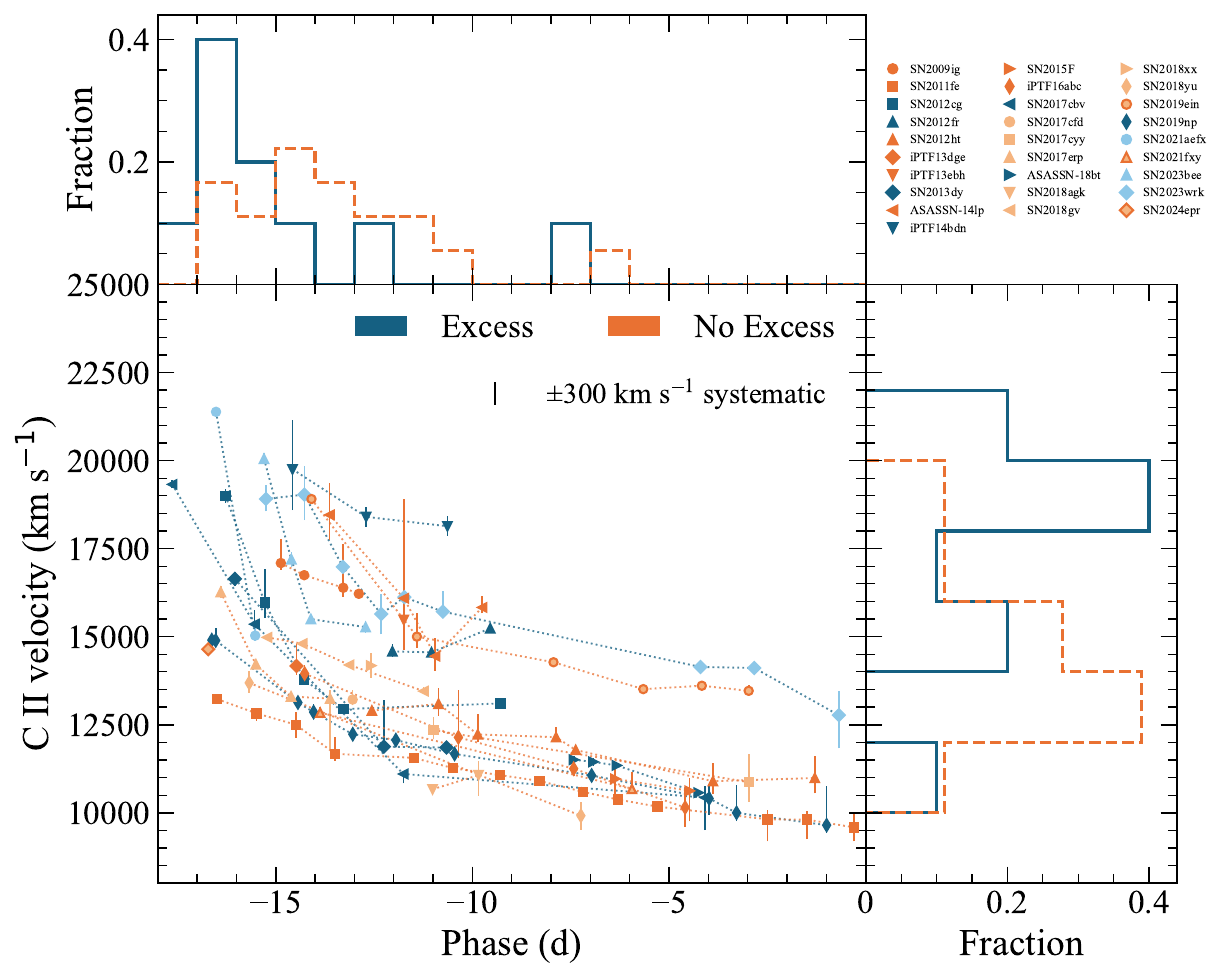}
\caption{Similar to Figure~\ref{fig:cpEW} but for \ion{C}{2} $\lambda6580$ velocity measurements. The black vertical error bar under the excess/no excess legend indicates an additional instrumental systematic uncertainty of $\pm300$~km~s$^{-1}$. The right histogram uses 2000~km~s$^{-1}$ bins.}
\label{fig:cvelocity}
\end{figure*}

\begin{figure*}[ht!]
\includegraphics[width=\textwidth]{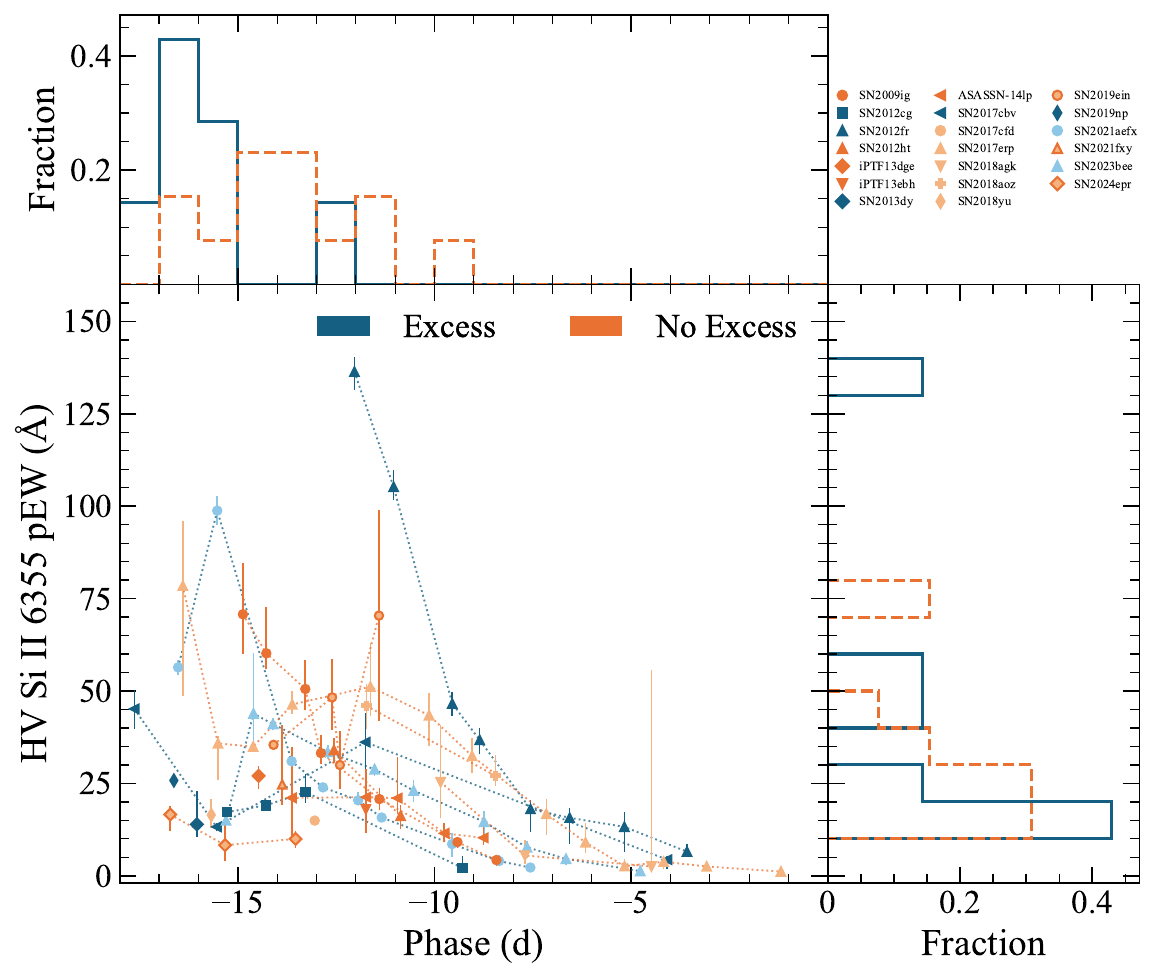}
\caption{Similar to Figure~\ref{fig:cpEW} but for HV \ion{Si}{2} $\lambda6355$ pEW measurements. The right histogram uses 10~\AA\ bins.}
\label{fig:hvsipew}
\end{figure*}

We examine the velocity evolution of \ion{C}{2} $\lambda6580$ in Figure~\ref{fig:cvelocity} to check the consistency of the carbon fits. For objects with multiple detections, velocities generally decrease with phase, consistent with the photosphere receding through the ejecta. A two-sample K--S test of the earliest available carbon velocity measurement for each object indicates a significant difference between the distributions ($D=0.600$, $p=0.00716$), with the flux-excess objects tending to have higher velocities. We interpret this with caution because the comparison depends strongly on phase coverage. Absorption velocities evolve rapidly before maximum light and generally decline more strongly with phase than pEW, so even small differences in the phase distribution can affect the apparent velocity separation. As shown in Figure~\ref{fig:cvelocity}, the largest velocity differences appear at the earliest phases, where the flux-excess objects are slightly overrepresented. Given these limitations, we treat this as an ancillary rather than a main finding of this work. As the sample of spectroscopically well-observed SNe\,Ia (i.e., with $\sim$daily cadence) with early excesses increases, future work may clarify the nature of this relationship. 

\begin{figure*}[t]
\includegraphics[width=\textwidth]{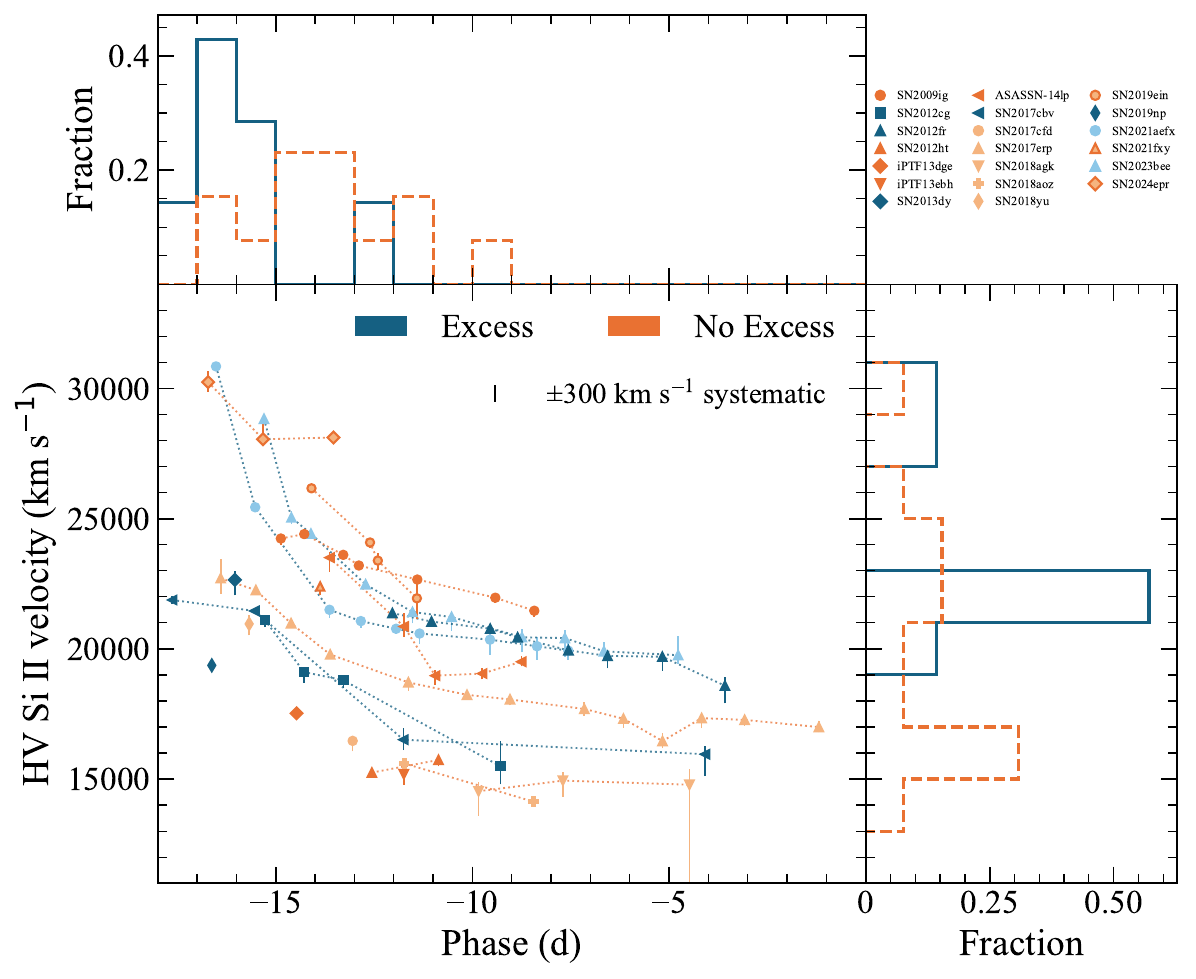}
\caption{Similar to Figure~\ref{fig:cpEW} but for HV \ion{Si}{2} $\lambda6355$ velocity measurements. The black error bar under the excess/no excess legend indicates an additional instrumental systematic uncertainty of $\pm300$~km~s$^{-1}$. The right histogram uses 2000~km~s$^{-1}$ bins.}
\label{fig:hvsivelocity}
\end{figure*}

Carbon non-detections among flux-excess SNe\,Ia require interpretation in the context of their early spectroscopic coverage and the rapid evolution of \ion{C}{2} absorption. The earliest spectra analyzed for LSQ12gdj, SNe\,2012fr, 2015bq, and 2021hpr occur at $-8.2$, $-12.0$, $-15.6$, and $-14.3$~d, respectively, and yield \ion{C}{2} pEW $<1$~\AA. For LSQ12gdj and SN\,2012fr, carbon absorption may have been present earlier but already weakened below detectability at these later phases.

SN\,2021hpr is a particularly interesting case because it is spectroscopically normal and has high-SNR spectra, yet all ten epochs from $-14.3$ to $-1.2$~d yield only $3\sigma$ upper limits on \ion{C}{2}. Nevertheless, the evolution of SN\,2021aefx illustrates why this does not exclude strong carbon absorption at earlier phases: the \ion{C}{2} pEW of SN\,2021aefx declines from $20.1$~\AA\ at $-16.5$~d to $6.0$~\AA\ at $-15.5$~d, reaching a $3\sigma$ upper limit of $<0.1$~\AA\ by $-13.6$~d. Thus, SN\,2021hpr could have declined rapidly, similar to SN~2021aefx, before the onset of spectroscopic coverage. Its low-resolution ($\sim18$~\AA) spectrum near $-13$~d also shows a possible flattened absorption feature redward of \ion{Si}{2}, but the identification of this as \ion{C}{2} remains ambiguous. A more constraining counterexample would therefore be a spectroscopically normal early excess with a high-SNR spectrum at $\lesssim-15$~d and a stringent non-detection of \ion{C}{2}.

\subsection{High-velocity Si}
\label{sec:hvsi}
A number of early spectra in our sample require an additional HV \ion{Si}{2} component to adequately fit the \ion{C}{2} $\lambda6580$ and \ion{Si}{2} $\lambda6355$ features, so we also examine the evolution of HV \ion{Si}{2} $\lambda6355$ across the sample. In Figures~\ref{fig:hvsipew} and \ref{fig:hvsivelocity}, HV \ion{Si}{2} pEWs and velocities do not clearly separate the two groups. The earliest HV \ion{Si}{2} pEWs have similar medians of 27 and 28 Å for early-excesses and non-excesses, respectively, while the corresponding median velocities are 21,600 and 20,900 km s$^{-1}$. When applying the same two-sample K--S test to the earliest HV \ion{Si}{2} measurements, we find no significant difference between the flux-excess and non-excess distributions at the $\alpha=0.05$ level, either in pEW ($D=0.286$, $p=0.734$) or velocity ($D=0.462$, $p=0.231$).

Several objects, including SNe\,2017cbv, 2019ein, and 2021aefx, show apparent increases during their HV \ion{Si}{2} pEW evolution. However, HVFs are expected to be strongest at early times and weaken rapidly as the photospheric component becomes dominant \citep[e.g.,][]{Mazzali05, Childress2014, Silverman15}. Interpreting this early evolution requires caution, since decomposing blended HV and photospheric \ion{Si}{2} absorption is sensitive to spectral cadence, SNR, and the assumed component structure. This is especially problematic when the transition from a HV-dominated profile to a photospheric-dominated profile is not well sampled (e.g., \citealp{Hosseinzadeh17}). Thus, the apparent unphysical evolution may reflect limitations in decomposing blended profiles.

\ion{C}{2} and prominent HV \ion{Si}{2} structure often coincide. Several flux-excess objects with \ion{C}{2} pEWs $>26$ \AA\ (e.g., SNe\,2012cg, 2017cbv, 2019np, and 2023bee) also show early \ion{Si}{2} profiles that are likely dominated by the HV component. 
Of the non-excesses with carbon detections, SNe\,2009ig, 2011fe, 2017erp, 2018yu, 2018agk, 2019ein, 2021fxy, and 2024epr show evidence for HV material. SN\,2023wrk, iPTF16abc, and iPTF14bdn show detectable early carbon and no clear HV \ion{Si}{2} structure, although they are 99aa-like objects. Such objects have weak early \ion{Si}{2} features, likely due to higher ionization \citep{Filippenko92, Phillips92, Garavini04, Phillips24}. This behavior is seen in both flux-excess and non-excess SNe\,Ia, suggesting that HV \ion{Si}{2} does not show the same association with early flux excesses as \ion{C}{2} absorption.

\begin{figure*}[ht!]
\includegraphics[width=\textwidth]{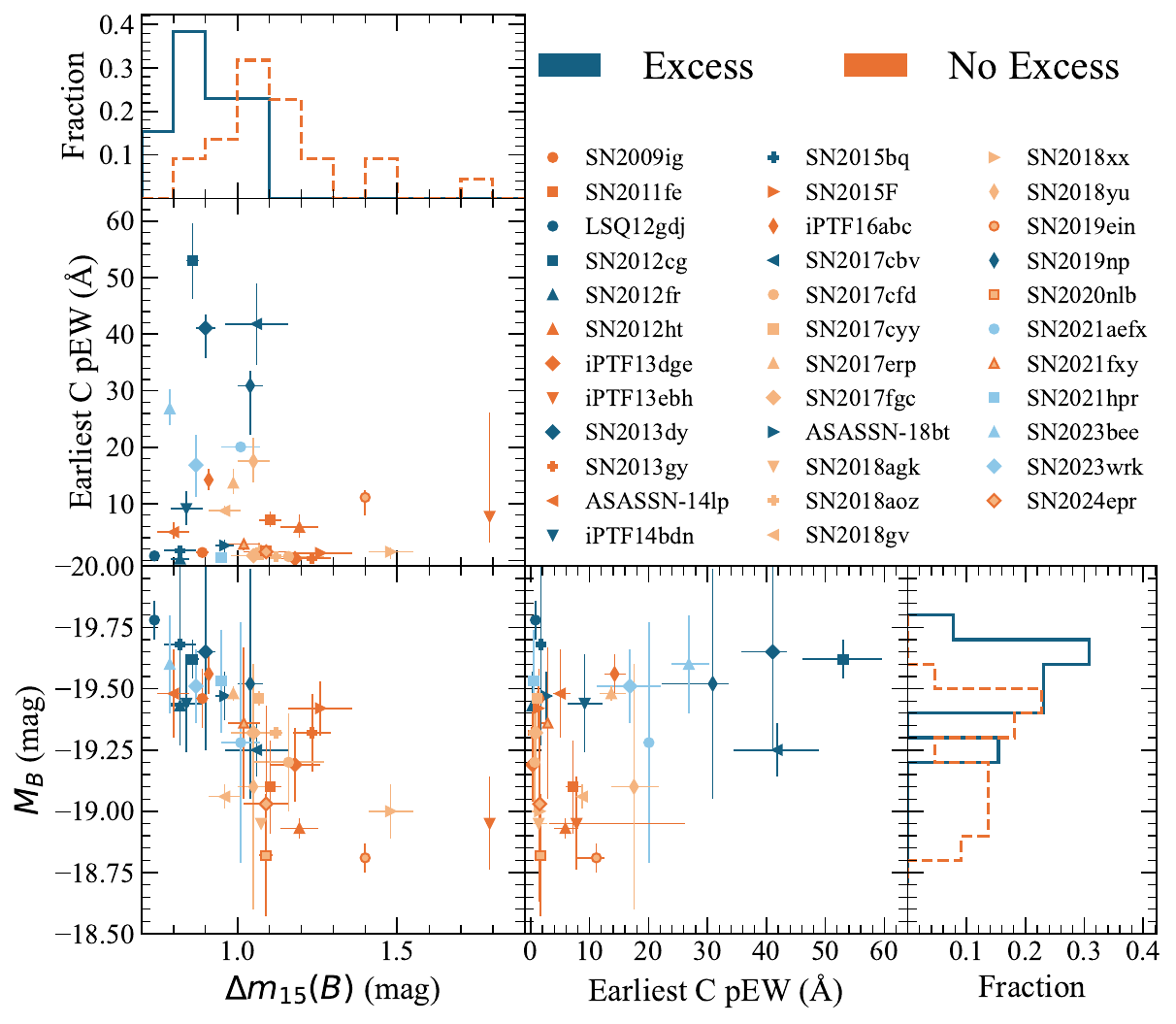}
\caption{Light-curve properties and earliest carbon absorption measurements. The bottom-left panel shows the relation between $\Delta m_{15}(B)$ and absolute $B$-band magnitude, $M_B$. The middle-left and bottom-middle panels compare the earliest measured \ion{C}{2} $\lambda6580$ pEW at phases $<0$ days relative to $B$-band maximum with $\Delta m_{15}(B)$ and $M_B$, respectively. The top-left and bottom-right panels show the distributions of $\Delta m_{15}(B)$ and $M_B$ for the flux-excess and non-excess subsamples, both using 0.1 mag bins. Blue and light blue points indicate SNe\,Ia with early flux excesses, while orange and light orange points indicate those without.}
\label{fig:lightcurve_properties}
\end{figure*}

\subsection{Sample Properties}
\label{sec:properties}
Differences in light-curve properties are significant because luminosity, decline rate, and spectroscopic subtype could instead reflect differences in the strength or detectability of \ion{C}{2}. Figure~\ref{fig:lightcurve_properties} shows the light-curve properties of the flux-excess and non-excess samples. The bottom-left panel shows the LWR for our sample, while the adjacent panels compare $\Delta m_{15}(B)$ and $M_B$ with the earliest \ion{C}{2} $\lambda6580$ pEW measurement for each object.

Objects with similar light-curve properties can show substantially different \ion{C}{2} strengths. For example, SN\,2015bq ($\Delta m_{15}(B)=0.82$ mag, $M_B=-19.68$ mag) has a $3\sigma$ upper limit of pEW $<2$~\AA\ at $-15.6$~d, whereas the similarly luminous and slowly declining SNe\,2012cg ($\Delta m_{15}(B)=0.86$ mag, $M_B=-19.62$ mag) and 2023bee ($\Delta m_{15}(B)=0.79$ mag, $M_B=-19.60$ mag) have early \ion{C}{2} pEWs of 54.9 and 28.1~\AA, respectively. Among the flux-excess objects, the weakest \ion{C}{2} measurements are concentrated among the few luminous, slowly declining 91T/99aa-like events, which may reflect a higher carbon ionization state. A systematic search for \ion{C}{3} in these luminous objects would confirm whether or not differences in light-curve properties contribute to the association between early excesses and stronger carbon absorption.

Our sample reproduces the observed tendency for flux-excess SNe\,Ia to be shifted toward luminous, slowly declining events. Previous studies have suggested that early excesses may be more common among 91T/99aa-like events \citep{Jiang18, Deckers2022}, which are broader and more luminous than normal SNe\,Ia \citep{Phillips24}. However, since these objects also tend to show stronger \ion{C}{2} absorption in our sample, this trend contrasts with previous studies that found carbon-positive SNe\,Ia have preferentially narrower light curves (e.g., \citealp{Thomas11, Maguire2014}). These studies sampled the broader SN\,Ia population and generally lacked the very early spectra that drive the separation in our sample. Even beyond location on the LWR, \citet{Ye24} find that early-excess SNe\,Ia have lower Hubble residuals at the $2–3\sigma$ level even after shape, color, and host-mass corrections.

\section{Discussion} \label{sec:discussion}
SNe\,Ia with early excesses show a population-level difference in \ion{C}{2} $\lambda6580$ strength, suggesting that early-excess SNe\,Ia differ systematically in their outer ejecta through ionization state, degree of burning, asymmetry/viewing angle, or the explosion model itself.

A variety of mechanisms have been proposed to explain early flux excesses in SNe\,Ia, which can broadly be organized by the source of the additional luminosity \citep[e.g.,][]{Jiang18}. These include ejecta-interaction with surrounding material (\S\ref{sec:interactions}) and radioactive heating from material in the outer ejecta (\S\ref{sec:radioactivity}). Previous studies have primarily evaluated these mechanisms through their ability to reproduce the early-time light curves, colors, and spectra of individual events. Here, we use the stronger \ion{C}{2} absorption observed in the flux-excess population as an additional constraint, assessing whether each mechanism can simultaneously account for the excess luminosity and the presence of unburned carbon in the outer ejecta. We summarize previous observational and theoretical work on each mechanism and, where appropriate, consider specific explosion or progenitor scenarios that provide physical realizations of them. We then discuss any possible discrepancies with ionization states in \S\ref{sec:ionizations}. Finally, in \S\ref{sec:hvsidisc}, we consider a potential association between HV \ion{Si}{2} and \ion{C}{2} and whether it could provide additional clues about the structure of the outer ejecta and the underlying explosion physics of SNe\,Ia.

\subsection{Interactions}\label{sec:interactions}
We first consider models in which the excess arises through interaction with a non-degenerate companion (\S\ref{sec:companion}), and then nearby circumstellar material (\S\ref{sec:CSM}). 

\subsubsection{Non-degenerate Companion}\label{sec:companion}
Shock interaction between the SN ejecta and a non-degenerate companion can produce an early UV/blue excess that is strongly dependent on binary separation and viewing angle \citep{Kasen10}. This scenario does not naturally predict a tendency for stronger \ion{C}{2} absorption in the SN ejecta. Furthermore, \citet{Jiang18} argued that viewing-angle dependence makes non-degenerate companion interaction unlikely to explain the high incidence of early excesses among luminous 91T/99aa-like SNe\,Ia, and \citet{Hoogendam24} extended this reasoning using 02es/03fg-like events, effectively eliminating the single-degenerate interaction scenario from consideration.
The early light curves and late-time limits on stripped H- or He-rich material further disfavor a dominant SD companion-interaction origin for most normal SNe\,Ia \citep{Leonard2007, Shappee13a, Shappee13b, Lundqvist15, Shappee18, Sand18, Botyanszki18, Tucker19, Tucker20, Fausnaugh21, Fausnaugh23, Tucker24}. Thus, while the SD scenario may remain relevant for individual oddballs and peculiar subtypes such as SNe\,Iax \citep{McCully14b, Stritzinger15, Jha17, Dubay22, Maguire23, Hoogendam25_24pxl}, it is likely not the source of an early-time excess flux.

\subsubsection{Circumstellar Material}\label{sec:CSM}
Interaction with a wind-like CSM can produce an early optical excess without necessarily producing the persistent signatures associated with SNe Ia-CSM \citep{Piro16, Moriya23}. Multiwavelength observations can constrain how prevalent this weaker interaction is among normal SNe\,Ia. Such signatures may be present at late times, but previous searches have yielded few detections among normal SNe\,Ia \citep{Graham19, Dubay22}, while radio surveys have generally resulted in non-detections \citep{Chomiuk16}. Deep X-ray non-detections of SNe\,2011fe and 2014J constrain the circumstellar environments expected from many single degenerate systems \citep{Margutti12, Margutti14}. These limits disfavor dense CSM around normal SNe\,Ia, but do not exclude weaker interaction capable of producing a short-lived early excess.

Interaction with C-rich material could connect early excesses and strong \ion{C}{2} absorption. DD scenarios provide a possibility, as violent WD mergers and WD collisions can leave substantial unburned material at high velocity or in an extended distribution surrounding the exploding WD \citep[e.g.,][]{Raskin13, Moll14, Raskin14, Heringer2019}. In these models, carbon absorption may arise from material originating in the secondary WD, particularly if it is disrupted or remains only partially burned during the explosion. In slow mergers, the lower-mass WD may be disrupted into a disk and gradually accreted by the primary \citep{Piersanti03}, where the resulting merger debris could form C-rich CSM. Interaction with nearby material could produce the early excess, while C-rich material associated with the merger could form the \ion{C}{2} feature. If this material is confined relatively close to the progenitor, it can be swept up at early times without requiring extended CSM that would produce detectable late-time UV or radio emission \citep{Chomiuk16, Graham19, Dubay22}.

The applicability of merger-produced CSM to normal SNe\,Ia remains unclear. Many existing envelope-interaction models have focused on peculiar or overluminous events \citep[e.g.,][]{Maeda23}, motivating investigations of whether DD merger debris can produce an early excess while maintaining the conditions required for detectable carbon absorption in otherwise normal SNe\,Ia.

\subsection{Surface Radioactivity}\label{sec:radioactivity}
We discuss mechanisms in which the excess is powered by radioactive material at shallow depths, distinguishing between $^{56}$Ni placed in the outer ejecta through explosion geometry (\S\ref{sec:shallowni}) and radioactive isotopes synthesized directly in a detonating He-shell (\S\ref{sec:shell}).

\subsubsection{Shallow $^{56}$Ni}\label{sec:shallowni}
When $^{56}$Ni is distributed into the outer ejecta, energy from its decay can diffuse out on a sufficiently short timescale to produce excess luminosity at early times \citep{Piro13, Piro16, Noebauer17, Jiang18}. Placing radioactive material in the outer ejecta will affect thermal and ionization structure, and iron-group elements associated with the radioactive material can increase UV line blanketing \citep{Piro16}. Sufficient heating is required to maintain singly ionized carbon, but stronger heating can shift an increasing fraction to \ion{C}{3} and weaken the \ion{C}{2} $\lambda6580$ feature \citep{Heringer2019}. Stronger \ion{C}{2} absorption in early excesses further constrains these models, which must produce enough shallow radioactive material to power an excess while preserving conditions that allow \ion{C}{2} to appear. 

Direct evidence that radioactive material can reach relatively shallow layers was provided by SN\,2014J. $^{56}$Ni $\gamma$-ray lines at 158 and 812~keV were detected $\sim$20~days after explosion \citep{Diehl14, Diehl15}, which \citet{Isern16} suggested to be a HV $^{56}$Ni plume. Importantly, the absence of detectable $^{56}$Ni $\gamma$-ray lines at even earlier times does not imply that shallow $^{56}$Ni was not affecting the light curve, as the outermost ejecta is initially opaque to $\gamma$-rays. SN\,2014J therefore provides evidence for possible radioactive material at high velocities, although the explosion physics responsible remain uncertain. 

Multi-dimensional explosion models show that delayed detonations can transport deflagration ashes at high velocities when plumes reach the WD surface before detonation \citep[e.g.,][]{Seitenzahl13}. For example, in the gravitationally confined detonation (GCD), an off-center deflagration reaches the surface and burned material sweeps around the star, converging on the opposite side where compression triggers a detonation \citep{Plewa04, Kasen05, Jordan08}. Early GCD models produced large $^{56}$Ni yields and were therefore most applicable to luminous SNe\,Ia \citep{Jordan08, Meakin09}. The resulting ejecta are highly asymmetric, with deflagration ashes extending to high velocities \citep{Kasen05, Meakin09}. More recent three-dimensional radiative-transfer calculations have shown that the high-velocity ashes can also produce viewing-angle-dependent light curves and spectra, including an early $U$-band excess for some lines of sight \citep{Seitenzahl16}. Whether GCD models can reproduce the diversity of early excesses remains to be explored in detail, particularly across different explosion configurations and viewing angles. 

Existing shallow-$^{56}$Ni calculations have provided detailed predictions for the early light curves, but they have not generally been used to test the \ion{C}{2} $\lambda6580$ absorption considered here. Radiative-transfer modeling shows that such distributions can reproduce aspects of the early light-curve morphology, although matching the accompanying color and spectral evolution remains difficult \citep{Dimitriadis19, Shappee19, Magee20a, Magee20b}. Investigating the evolution of \ion{C}{2} within these models in the first few days after explosion would directly test whether shallow radioactive material can simultaneously reproduce the early excess and the stronger carbon absorption observed in the flux-excess population.

\subsubsection{He-shell detonations}\label{sec:shell}
He-shell double-detonations provide a physically distinct route to form radioactive material in the outer ejecta. In these models, a surface He-shell detonation produces radioactive isotopes such as $^{56}$Ni, $^{52}$Fe, and $^{48}$Cr before triggering a detonation in the underlying CO core. The radioactive decay could increase the luminosity at early times, with the resulting excess depending on the shell mass and composition \citep{Polin19, Perets2019, Magee21}. Configurations invoked to explain early excesses among normal SNe\,Ia must also account for increased \ion{C}{2} absorption.

The composition of the shell-burning products strongly influences whether He-shell double-detonation models can resemble normal SNe\,Ia. Massive He-shells can produce substantial amounts of iron-group and intermediate-mass elements in the outer ejecta, leading to strong UV/blue line blanketing and colors inconsistent with those of normal SNe\,Ia. Thinner shells can instead produce spectra with less line blanketing and colors more consistent with normal SNe\,Ia \citep[e.g.,][]{Townsley19, Boos21, Shen21, Boos24b}. Multidimensional models introduce further diversity through asymmetric distributions of shell-burning products and viewing-angle-dependent spectra \citep[e.g.,][]{Shen21}.

\begin{figure}[t!]
\centering
\includegraphics[width=\columnwidth]{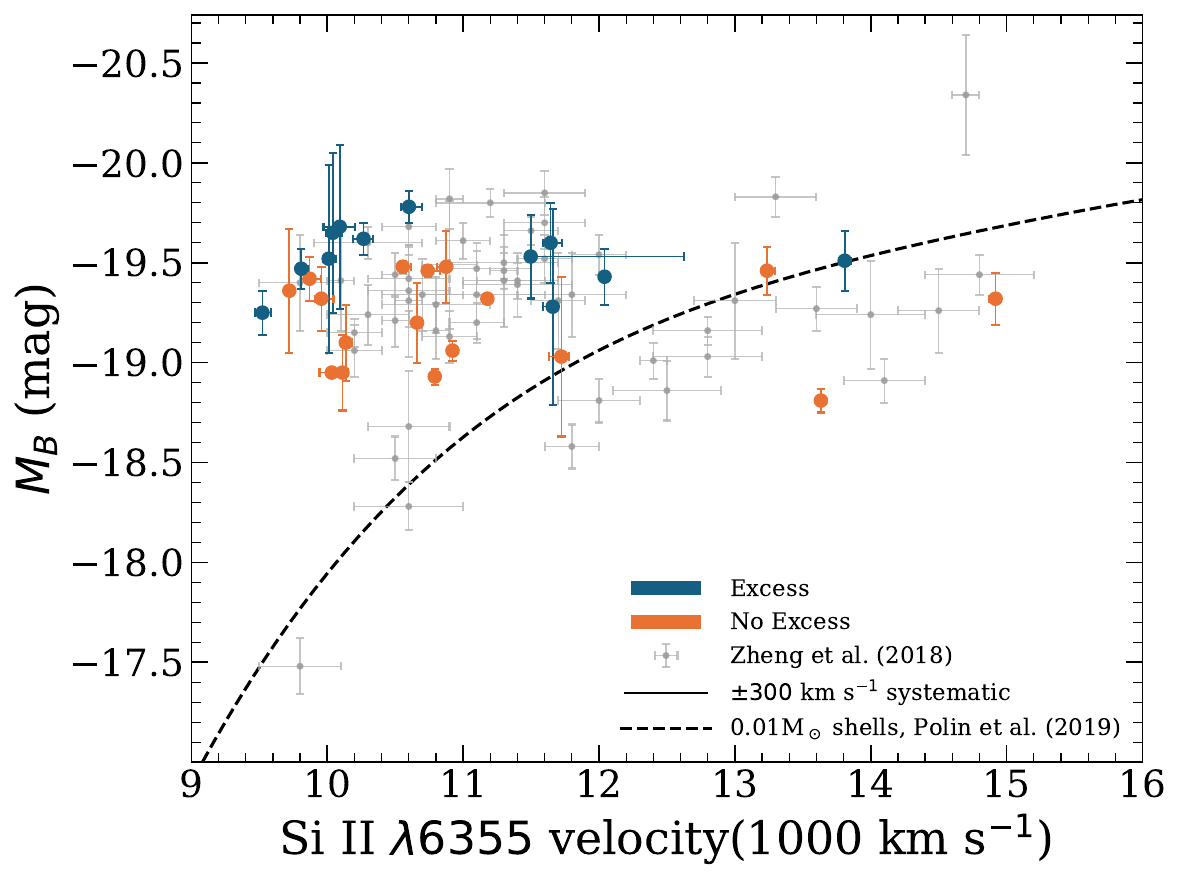}
\caption{$M_B$ versus \ion{Si}{2} $\lambda6355$ velocity for our sample. Velocities are measured from the spectrum nearest to maximum light in the $B$-band ($\pm3$ days). The dashed line shows the theoretical relationship predicted by 1D double-detonation models from \citet{Polin19}, with helium shells of mass 0.01M$_\odot$ and varying progenitor masses. The gray points are a sample taken from \citet{Zheng18}.}
\label{fig:sivelocity}
\end{figure}

Figure~\ref{fig:sivelocity} compares the peak absolute $B$-band magnitudes and near-maximum \ion{Si}{2} $\lambda6355$ velocities of our sample with the thin He-shell double-detonations from \citet{Polin19} models\footnote{Accessed via \url{https://github.com/slowdivePTG/SN2022joj/blob/main/photometry/phase_space.ipynb}}. We include SNe\,Ia with spectra $\pm3$~d from maximum light.

A potential DD pathway for He-shell detonations is the dynamically driven double degenerate double-detonation (D$^6$). In this picture, the shell is formed with He-rich mass transfer from a He WD \citep{Guillochon10, Shen18}. Synthetic light curves and spectra from these models show that they may produce early UV/blue excesses for some parameters and can generate asymmetric, viewing-angle-dependent outer ejecta with possible carbon absorption \citep{Shen21}. More recent work has explored whether the donor may also detonate, which would further modify the ejecta geometry and observable signatures \citep{Pakmor22, Boos24a}.

Regardless of the progenitor pathway that produces the double detonation, its radioactive burning products provide a potential observational test of the mechanism. In addition to $^{56}$Ni, the decay chains of $^{52}$Fe and $^{48}$Cr can contribute to the early luminosity \citep{Magee21}. The upcoming Compton Spectrometer and Imager (COSI) will provide improved sensitivity to MeV $\gamma$-ray lines \citep{Tomsick23}, which could constrain the amount of shallow $^{56}$Ni expected in He-shell detonation models.

\subsection{Carbon Ionization States}\label{sec:ionizations}
Differences in ionization state could explain the observed \ion{C}{2} preference. As SNe\,Ia decrease in luminosity and cool, carbon may recombine from \ion{C}{2} to \ion{C}{1}, and some SNe\,Ia could have low \ion{C}{2} pEW values but still be C-rich. The ratio of the absorption depths of \ion{Si}{2} $\lambda5972$ to $\lambda6355$, $R(\mathrm{Si~II})$, is sensitive to the ionization state and provides a proxy for photospheric temperature near maximum light \citep{Nugent95, Hachinger08}.

\begin{figure}[t!]
\centering
\includegraphics[width=\columnwidth]{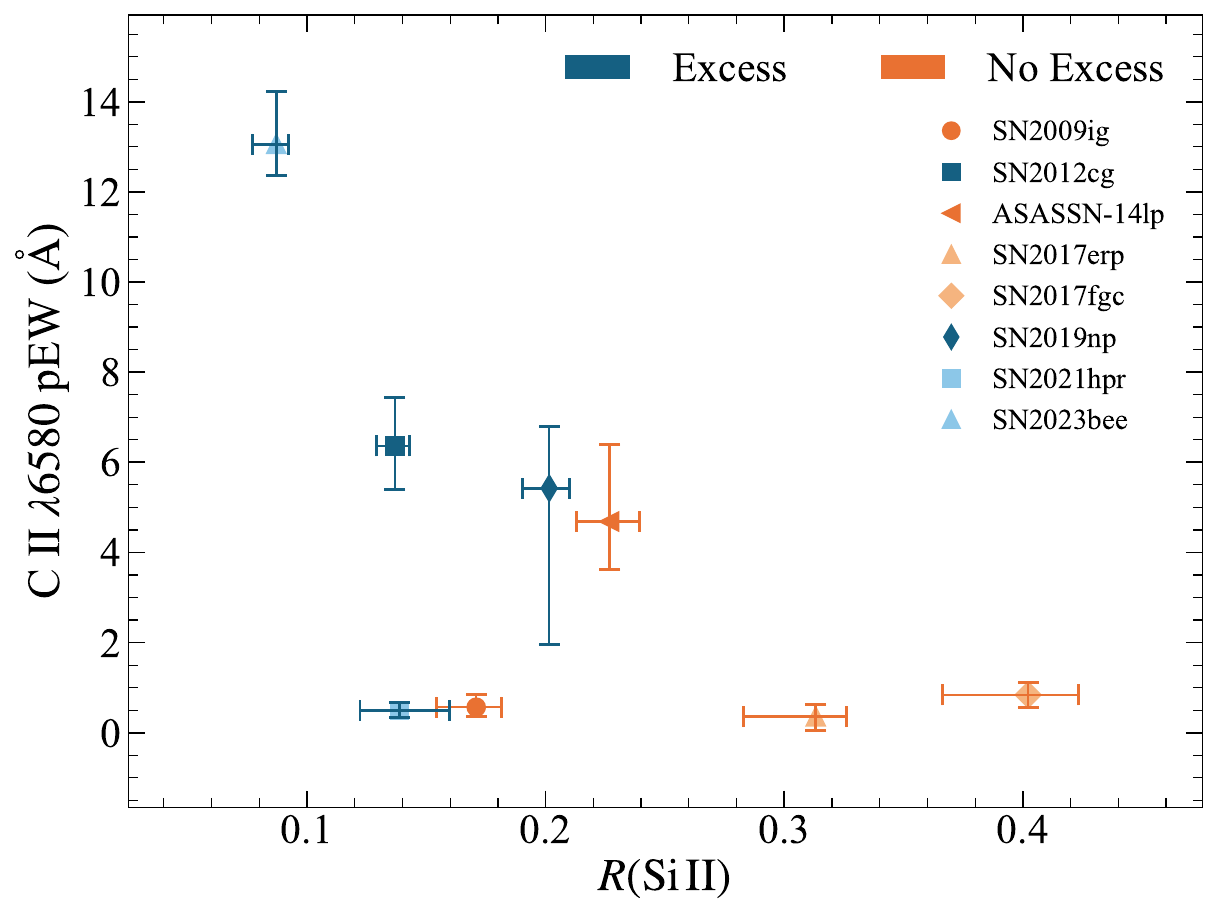}
\caption{\ion{C}{2}~$\lambda6580$ pEW versus $R(\mathrm{Si\,II})$ for spectra obtained between $-14.5$ and $-13.5$~d relative to $B$-band maximum. Blue and orange symbols denote flux-excess and non-excess SNe\,Ia, respectively, with individual objects identified by their marker styles.}
\label{fig:rsi}
\end{figure}

Figure~\ref{fig:rsi} compares \ion{C}{2} $\lambda6580$ and $R(\mathrm{Si~II})$ for SNe\,Ia with spectra obtained between $-14.5$ and $-13.5$~d. We restrict the sample to objects with similar maximum-light properties, spanning $-19.62 \leq M_B \leq -19.32$~mag and $0.79 \leq \Delta m_{15}(B) \leq 1.05$~mag. A particularly useful comparison is SN\,2012cg and SN\,2009ig at $-14.3$~d. Their silicon depth ratios are $0.14^{+0.01}_{-0.01}$ and $0.17^{+0.01}_{-0.02}$, respectively, whereas their \ion{C}{2} pEWs differ by an order of magnitude: $6.4^{+1.1}_{-1.0}$~\AA\ versus $0.6^{+0.3}_{-0.2}$~\AA. Furthermore, SN\,2019np at $-14.0$~d has $R(\mathrm{Si~II})=0.20^{+0.01}_{-0.01}$ and a \ion{C}{2} pEW of $5.4^{+1.4}_{-3.5}$~\AA. This shows that, at least in some cases, when objects have similar light-curve shape and photospheric temperature, the effect we see appears not to be driven by ionization but rather by distinct differences in the explosion, such as the abundance and density of carbon, and the efficiency of the burning in the WD.

For the most luminous SNe\,Ia, weak \ion{C}{2} absorption is difficult to interpret because the presence of carbon in higher ionization states remains uncertain. iPTF14bdn, iPTF16abc, and SN2023wrk show higher-ionization features consistent with their spectroscopic subtyping as a 99aa-like SN\,Ia \citep{Scalzo14, Li22}. These objects have sufficient temperature to ionize \ion{Si}{2} to \ion{Si}{3}, resulting in weak or absent \ion{Si}{2} features throughout their photospheric spectra \citep{Phillips24}, lacking a quantitative $R(\mathrm{Si~II})$. However, the presence of \ion{Si}{3} does not establish that carbon is ionized to \ion{C}{3}. The threshold from \ion{Si}{2} to \ion{Si}{3} is 16.35~eV, compared with 24.38~eV for \ion{C}{2} to \ion{C}{3}, allowing \ion{Si}{3} to coexist with \ion{C}{2}. However, spectral modeling of LSQ12gdj suggests possible \ion{C}{3} $\lambda4649$ despite no evident \ion{C}{2} $\lambda6580$ absorption \citep{Scalzo14}, so higher ionization remains a possible explanation for weak \ion{C}{2} in 91T/99aa-like objects.

\subsection{High-velocity \ion{Si}{2}}\label{sec:hvsidisc}
The coexistence of detectable \ion{C}{2} and HV \ion{Si}{2} in both light-curve groups motivates a closer examination of their relationship. \citet{Silverman15} reported that HV \ion{Si}{2} tends to occur in objects without detected \ion{C}{2}, contrasting with the coexistence observed in several objects in our sample. Differences in spectral coverage may contribute to this discrepancy, while uncertainty in the \ion{Si}{2} profile decomposition further complicates the comparison. The strongest HV \ion{Si}{2} pEWs occur in objects whose early \ion{Si}{2} $\lambda6355$ profiles appear to be dominated by the HV component (e.g., \citealp{Childress13, Hosseinzadeh17}). Limited cadence can make it difficult to follow the transition to a photospheric-dominated profile and assign the fitted components consistently across epochs. In addition, some objects (e.g., SNe\,2011fe and 2013dy) with ambiguous HV \ion{Si}{2} show evidence for other HVFs \citep{Pereira13, Zhang16_11fe, Zhai16}.

The physical origin of HV \ion{Si}{2} remains uncertain, further limiting interpretation. HVF strength depends on the ejecta's density, ionization, and excitation. For example, HV \ion{Ca}{2} is observed substantially more often than HV \ion{Si}{2}, in part because the strong \ion{Ca}{2} transitions have lower excitation requirements \citep{Mazzali05, Maguire2014, Silverman15}. Recent radiative-transfer calculations show that HV density enhancements can reproduce much of the observed HV \ion{Si}{2} evolution \citep{Harvey26}, although current delayed-detonation and double-detonation models do not reproduce the full diversity of observed HVFs. 

Higher SNR spectra with moderate resolving power ($R\sim1000$) before $\sim -14$~d are needed to establish how \ion{C}{2} and HV \ion{Si}{2} evolve relative to one another and to better separate HV and photospheric \ion{Si}{2} components. Nebular-phase line profiles offer a complementary probe of asymmetries in the inner ejecta \citep[e.g.,][]{Vallely20}, so comparing these profiles with early \ion{C}{2}, HV \ion{Si}{2}, and flux-excess behavior could help determine whether the early-time diversity is related to viewing angle.

\section{Conclusion} \label{sec:conclusion}
We investigate early-time spectra of SNe\,Ia with well-observed rising light curves, measuring pre-maximum \ion{C}{2} $\lambda6580$ absorption and the blended \ion{Si}{2} $\lambda6355$ profile. SNe\,Ia with early excesses are preferentially brighter and show systematically stronger \ion{C}{2} absorption at the earliest phases than the non-excesses, while the differences in pEWs and velocities of the HV \ion{Si}{2} components are not statistically significant. A two-sample K--S test rejects the hypothesis that the earliest \ion{C}{2} pEW measurements of the two populations are drawn from the same parent distribution at the 5\% significance level ($p=0.025$), suggesting that early excesses are associated with differences in the composition or ionization of unburned material in the outer ejecta.

This correlation provides a new constraint on the physical origin of early excesses.~CSM interaction remains possible if the material is C-rich.~Models that place radioactive material in the outer ejecta, including shallow-$^{56}$Ni and He-shell detonation scenarios, must produce the early luminosity without over-ionizing, burning, or obscuring the observable \ion{C}{2}. Both $M_{\text{Ch}}$ and sub-$M_{\text{Ch}}$ may produce radioactive material at high velocity while also generating strongly structured outer ejecta in which unburned carbon may remain observable. These scenarios offer avenues to test whether the same explosion physics can explain both the early excesses and the stronger carbon absorption. The observed correlation does not, however, uniquely identify a progenitor system or explosion mechanism, particularly because multidimensional structure and viewing angle can affect both early luminosity and carbon detectability.

However, HV \ion{Si}{2} is common in both flux-excess and non-excess SNe\,Ia at early times and does not show the same population-level association with early excesses as \ion{C}{2}. Nevertheless, its physical connection to unburned material and early-time behavior remains an important topic for future study.

Individual SNe\,Ia can be strongly affected by viewing angle, phase coverage, and object-specific peculiarities, whereas the systematic difference found here links early light-curve behavior to outer-ejecta composition across a sample of events. More spectra obtained within the first few days after explosion, along with high-cadence UV/optical photometry, will allow for the direct comparison of \ion{C}{2} strength with the amplitude and morphology of early excesses. Detailed radiative-transfer calculations should simulate combined distributions of radioactive material, unburned carbon, and HVFs. Together, these observations and models can establish the origin of early-excess SNe\,Ia and whether they represent a distinct subset of explosions or are part of a continuum within the broader SN\,Ia population.

\begin{acknowledgments}

We thank Tony Piro and Michael Tucker for helpful suggestions and discussions, and we appreciate the constructive conversations at the ``Type-Ia Supernova Progenitors and Cosmology in Light of DESI'' workshop hosted at the Ohio State University Center for Cosmology and AstroParticle Physics (CCAPP).

L.P.D. acknowledges support from NSF through the University of Hawai\okina i at M\={a}noa Institute for Astronomy REU program under grant No. 2548017, and would like to thank the Institute for Astronomy for a rewarding experience.

W.B.H. acknowledges support from the NSF Graduate Research Fellowship. This material is based upon work supported by the National Science Foundation Graduate Research Fellowship Program under Grant Nos. 1842402 and 2236415. Any opinions, findings, conclusions, or recommendations expressed in this material are those of the author(s) and do not necessarily reflect the views of the National Science Foundation.

This research made use of the Weizmann Interactive Supernova Data Repository (WISeREP; \citealp{Yaron2012}) and NASA's Astrophysics Data System.

\software{
Astropy \citep{2013A&A...558A..33A, 2018AJ....156..123A, 2022ApJ...935..167A},
Matplotlib \citep{matplotlib},
NumPy \citep{numpy},
pandas \citep{pandas},
SciPy \citep{scipy},
SpectRes \citep{Carnall2017}
}

\end{acknowledgments}

\begin{figure*}[htbp]
    \centering
    \includegraphics[width=\textwidth]{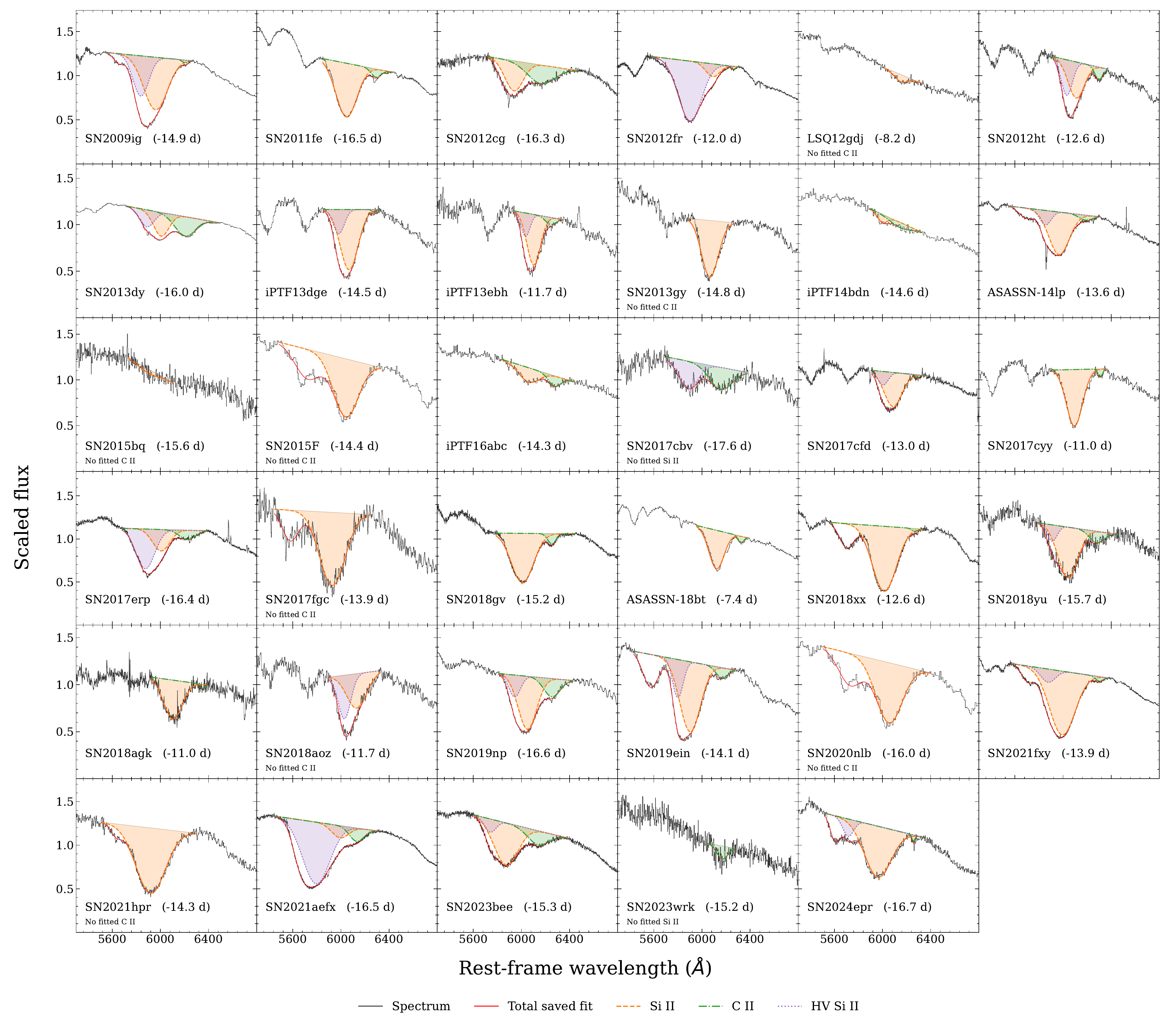}
    \caption{Earliest spectral measurements of the 35 SNe\,Ia in our sample. Each spectrum is scaled by its median flux between 5200 and 7000 \AA\.~Object names and phases relative to $B$-band maximum are indicated in each panel. The red curves show the combined fit, while the orange dashed, green dash-dotted, and purple dotted curves show the individual \ion{Si}{2}, \ion{C}{2}, and HV \ion{Si}{2} components, respectively. Shaded regions highlight their absorption profile.}
    \label{fig:spectralgrid}
\end{figure*}

\begin{appendix}
We present the spectral fits of the silicon and carbon profiles for the earliest measurements of our 35 SNe\,Ia in Figure~\ref{fig:spectralgrid}.

\end{appendix}

\bibliography{SNeIa}{}
\bibliographystyle{aasjournal}

\end{document}

%% file: tables/measurements_table_preview.tex
\begin{table*}[ht]
\centering
\setlength{\tabcolsep}{3pt}
\renewcommand{\arraystretch}{1.25}
\caption{Spectroscopic measurements for the \ion{C}{2} $\lambda6580$, \ion{Si}{2} $\lambda6355$, and high-velocity \ion{Si}{2} $\lambda6355$ absorption features. Phase is calculated relative to $B$-band maximum light in the rest frame. Positive and negative uncertainties are listed in separate columns. When carbon is undetectable, C pEW is computed from the $3\sigma$ upper-bound with a corresponding $1\sigma$ uncertainty using $W = 50$ for the C II Gaussian (equation~\ref{eq:upperlimit}). The full table is available in machine-readable form as an ancillary file.}
\label{tab:measurements}
\begin{tabular}{llcccccccc}
\hline
\shortstack{\strut Object} & \shortstack{\strut Component} & \shortstack{\strut MJD} & \shortstack{\strut Phase \\ (days)} & \shortstack{\strut pEW \\ (Å)} & \shortstack{\strut $+\sigma_{\rm pEW}$ \\ (Å)} & \shortstack{\strut $-\sigma_{\rm pEW}$ \\ (Å)} & \shortstack{\strut Velocity \\ ($\mathrm{km~s^{-1}}$)} & \shortstack{\strut $+\sigma_v$ \\ ($\mathrm{km~s^{-1}}$)} & \shortstack{\strut $-\sigma_v$ \\ ($\mathrm{km~s^{-1}}$)} \\
\hline
SN\,2009ig & C & $55065.0$ & $-14.9$ & $1.080$ & $0.528$ & $0.438$ & $17090$ & $691$ & $192$ \\
SN\,2009ig & Si & $55065.0$ & $-14.9$ & $108.378$ & $12.096$ & $15.617$ & $18207$ & $405$ & $543$ \\
SN\,2009ig & HV Si & $55065.0$ & $-14.9$ & $70.755$ & $13.760$ & $10.636$ & $24240$ & $144$ & $262$ \\
SN\,2009ig & C & $55065.6$ & $-14.3$ & $0.539$ & $0.181$ & $0.321$ & $16746$ & $75$ & $66$ \\
SN\,2009ig & Si & $55065.6$ & $-14.3$ & $123.833$ & $5.583$ & $15.433$ & $18652$ & $156$ & $458$ \\
SN\,2009ig & HV Si & $55065.6$ & $-14.3$ & $60.203$ & $12.467$ & $4.265$ & $24410$ & $46$ & $151$ \\
SN\,2009ig & C & $55066.6$ & $-13.3$ & $0.116$ & $0.178$ & $0.083$ & $16388$ & $1256$ & $256$ \\
SN\,2009ig & Si & $55066.6$ & $-13.3$ & $115.372$ & $7.386$ & $10.766$ & $17663$ & $216$ & $312$ \\
SN\,2009ig & HV Si & $55066.6$ & $-13.3$ & $50.550$ & $7.933$ & $5.615$ & $23613$ & $75$ & $139$ \\
SN\,2009ig & C & $55067.0$ & $-12.9$ & $0.079$ & $0.053$ & $0.062$ & $16216$ & $61$ & $66$ \\
\ldots & \ldots & \ldots & \ldots & \ldots & \ldots & \ldots & \ldots & \ldots & \ldots \\
SN\,2024epr & C & $60391.2$ & $-15.3$ & $0.967$ & $0.322$ & $0.322$ & \nodata & \nodata & \nodata \\
SN\,2024epr & Si & $60391.2$ & $-15.3$ & $144.051$ & $8.015$ & $7.375$ & $17135$ & $71$ & $84$ \\
SN\,2024epr & HV Si & $60391.2$ & $-15.3$ & $8.286$ & $8.559$ & $4.192$ & $28052$ & $758$ & $237$ \\
SN\,2024epr & C & $60393.0$ & $-13.5$ & $0.636$ & $0.212$ & $0.212$ & \nodata & \nodata & \nodata \\
SN\,2024epr & Si & $60393.0$ & $-13.5$ & $151.453$ & $5.005$ & $4.431$ & $16429$ & $49$ & $48$ \\
SN\,2024epr & HV Si & $60393.0$ & $-13.5$ & $9.984$ & $2.046$ & $2.445$ & $28119$ & $249$ & $228$ \\
SN\,2024epr & C & $60396.2$ & $-10.3$ & $0.541$ & $0.180$ & $0.180$ & \nodata & \nodata & \nodata \\
SN\,2024epr & Si & $60396.2$ & $-10.3$ & $154.571$ & $3.461$ & $4.696$ & $14500$ & $45$ & $50$ \\
SN\,2024epr & C & $60406.2$ & $-0.4$ & $0.685$ & $0.228$ & $0.228$ & \nodata & \nodata & \nodata \\
SN\,2024epr & Si & $60406.2$ & $-0.4$ & $171.854$ & $3.765$ & $8.888$ & $11723$ & $55$ & $92$ \\
\hline
\end{tabular}
\end{table*}